\documentclass[%
reprint,onecolumn,notitlepage,
 amsmath,amssymb,
longbibliography
]{revtex4-1}

\usepackage{subfig}
\usepackage{mathbbol}
\usepackage[nameinlink,capitalise,noabbrev]{cleveref}
\usepackage{xcolor}
\usepackage{pgfplots}
\usepackage{tikz}
\usetikzlibrary{arrows.meta,arrows,patterns}
\usetikzlibrary{positioning,shapes}
\usetikzlibrary{plotmarks}
\usetikzlibrary{mindmap}
\usetikzlibrary{spy}
\usepackage{xparse}
\usepackage[labelfont=bf]{caption}

\graphicspath{{./}{./images/}}

\NewDocumentCommand\dsum{e{_^}}{{\displaystyle\sum_{#1}^{#2}}}

\DeclareSymbolFontAlphabet{\amsmathbb}{AMSb}%

\begin{document}

\preprint{APS/123-QED}

\title{Dense bidisperse suspensions under non-homogeneous shear}

\author{Alessandro Monti (alessandro.monti@oist.jp)}
\author{Marco Edoardo Rosti (marco.rosti@oist.jp)}
\affiliation{Complex Fluids and Flows Unit, Okinawa Institute of Science and Technology Graduate University, 1919-1 Tancha, Onna-son, Okinawa 904-0495, Japan}%

\date{\today}

\begin{abstract}
We study the rheological behaviour of bidisperse suspensions in three dimensions under a non-uniform shear flow,
made by the superimposition of a linear shear and a sinusoidal disturbance.
Our results show that \textit{i)} only a streamwise disturbance in the shear-plane alters the
suspension dynamics by substantially reducing the relative viscosity, 
\textit{ii)} with the amplitude of the
disturbance determining a threshold value for the effect to kick-in and its wavenumber
controlling the amount of reduction and which of the two phases is affected. 
We show that,  \textit{iii)} the rheological changes are caused by
the effective separation of the two phases, with the large or small particles layering in separate regions.
We provide a physical explanation of the phase separation process
and of the conditions necessary to trigger it. We test the results in the whole flow curve,
and we show that the mechanism remains substantially unaltered, with the only 
difference being the nature of the interactions between particles modified
by the phase separation.
\end{abstract}

\maketitle

\section{Introduction}
Dense suspensions of hard particles immersed in a Newtonian solvent 
are common in many natural and industrial applications, such as
the pharmaceutical industry (e.g.\ the process of blending powders
for tablets production), food industry (e.g.\ powdery products),
powder metallurgy (e.g.\ the processes of compacting and sintering
blended pulverized metals), and sediment transport (e.g. transportation 
of nutrients and landscape shaping) \citep{MUZZIO2003,FITZPATRICK2005,%
VERCRUYSSE2017,DEGARMO1997,GILLISSEN2020}, with applications reaching
biofluidics and bacterial suspensions \citep{GUO2018}. 
These materials, when subject to a shear-rate $\dot{\gamma}$, often show 
peculiar rheological behaviours that can increase or decrease the fluidity 
of the suspension. Among these behaviours, shear-thickening, i.e.\ a
reduction of the fluidity properties of the suspension when the latter 
is subject to an increasing $\dot{\gamma}$, is of particular interest. 
Shear thickening is perhaps the most astonishing and most studied
non-Newtonian behaviour of dense suspensions, and until few years ago
it was far from being understood. 
The reduction of the fluidity is now attributed to the modification of the 
internal microstructure of the fluid \citep{STICKEL2005}; a recent 
theory experimentally proved that shear-thickening is triggered by the 
close interactions between the particles that 
appear in the form of frictional contacts (due to the microscopic asperities on the 
surface of the particles) and constrain their relative movements,
causing an enhancement of the viscosity of the suspensions 
\citep{BEHRINGER2008, SETO2013, FERNANDEZ2013, MARI2014, WYART2014, LIN2015, MARI2015, THOMAS2018, SINGH2019, SINGH2020, RATHEE2021}.
The intensity of the shear-thickening can be quantified as an increase of 
the effective viscosity of the suspension $\eta_r$ and becomes stronger 
as the volume fraction of the suspension $\phi$ increases,
until it displays an abrupt enhancement, behaviour known as discontinuous
shear thickening \citep{MORRIS2020, THOMAS2020}.
The dependence on $\dot{\gamma}$ and $\phi$ of the rheological 
properties generally characterizes the mechanics of the suspensions within the 
shear-thickening regime.
When the particles are suspended in a simple homogeneous shear flow, 
\citet{BOYER2011} demonstrated that such dual dependence can be 
reduced to a single parameter in analogy to granular flows \citep{JOP2006}. 
In fact, specifying the macroscopic friction coefficient
$\mu=\sigma_{12}/\Pi$ uniquely sets the volume fraction and the shear-rate,
and general constitutive laws $\eta_r=\eta_r(\mu)$ and $\phi=\phi(\mu)$ 
can be formulated \citep{BOYER2011, GILLISSEN2020PRL}. 
In the previous relation, $\sigma_{12}$ is the shear component of the stress tensor
$\sigma$ and $\Pi$ is the pressure, related to the
trace of $\sigma$. 
This powerful general outcome, however, fails
when dealing with spatial or temporal inhomogeneity within the flow. 
In fact, phenomena such as subyielding and overcompaction 
\citep{NOTT1994, HAMPTON1997, OH2015}, particles migration \citep{MATAS2004,
FAN2014, ITOH2019} or flow instabilities that can lead to
banding \citep{BESSELING2010, CHACKO2018, SAINTMICHEL2018, DEVITA2020} and 
segregative phenomena \citep{OTTINO2000} cannot be captured by such
constitutive-law, since it would require a two-phase description
\citep{NESS2022, GILLISSEN2020PRL}. The route towards a complete model
able to describe real suspensions that involve the aforementioned phenomena
requires more studies that shed some light onto the elusive physical 
mechanisms activated by the inhomogeneities.

In this direction, we propose a three-dimensional numerical study that 
aims at analysing non-trivial phenomena generated when bidisperse particles 
are suspended in an inhomogeneous Newtonian flow 
characterized
by the linear super-imposition of a plain shear flow and a sinusoidal velocity 
profile, i.e.,~ with zero mean but a non-uniform shear-rate.
Similar numerical experiments were carried out on two-dimensional granular flows:
e.g.,~\citet{SAITOH2019} studied a Kolmogorov-like flow of soft athermal
disks close to the jamming transition and found that stress profiles can be
reproduced by nonlocal constitutive relations that account for
fourth-order gradients \citep[see also][]{BAUMGARTEN2017}.

The importance of studying non-uniform profiles is suggested e.g.,~by the 
experimental shear-rheometer, where the condition of plain shear is 
hard to reach due the presence of the solid walls that inevitably introduce
local inhomogeneity. Also, the present configuration is a quite general
configuration of any pressure driven flows in channels, which are characterised
by a non-uniform mean shear-rate \citep{MUSACCHIO2014}.
In particular, the goal of the work is to trigger a mechanism of
particles migration that results in a demixing of the two phases. In this 
sense, many works dealt with dense granular flow, where demixing 
effects (e.g.,\ segregation) are of paramount interest to control and 
improve the industrial products quality
\citep{SAVAGE1993a, SAVAGE1993b, DRAHUN1983, OTTINO2000, WIEDERSEINER2011,
TUNUGUNTLA2014, UMBANHOWAR2019, DUAN2020, DUAN2021}.
To unravel the physical mechanisms governing the segregation, many recent 
studies focused on the the single intruder particle limit \citep{TRIPATHI2013,
GUILLARD2016, JING2017, VANDERVAART2018, JING2020, TREWHELA2021, JING2021}
since it provides a simple, yet effective, test-case to investigate the problem.
In particular, \citet{GUILLARD2016} carried out 2D numerical simulations
where they proposed a method to measure the force acting on the intruder.
This method was based on a virtual spring attached to the intruder that 
constrained it through a restoring force to oscillate around an equilibrium position. 
In this way, the authors were able to measure the force acting on the intruder and 
to derive an expression that modelled the segregative force as a function of the 
local pressure and the local shear-stress gradients. Following this seminal work,
\citet{VANDERVAART2018, JING2020} showed that the segregation force is insensitive
to the shear-stress gradients when the granular flow is subject to a linear velocity
profile (typical of free-surface flows). However, when the velocity profile is 
non-linear, a higher-order correction that takes into account the local shear-rate 
gradients has to be included to correctly model the segregative force \citep{JING2021}.
While these results have a great potential outcome to reach a unified continuum
model, the studies are actually limited to a single intruder particle, and the
origin of the segregation is still unclear. Finally, these models are limited
to granular flows, where only contacts forces acting between the particles are
modelled. A few studies on dilute and semi-dense suspensions 
are available in the literature \citep{KRISHNAN1995,KRISHNAN1996,ZARRAGA2001}, where
the authors mainly focused on the diffusivity of the suspensions and on some aspects
of the shear-induced segregation.

In this work, we tackle these open-questions via a three-dimensional numerical investigation 
that studies the rheology of a dense bidisperse suspension, with high dispersion 
ratio \citep{MONTI2023},
driven by a linear combination of a plain shear-flow and a sinusoidal disturbance. We will exploit the numerical techniques to study the effect of the amplitude and of the wavenumber of the disturbance
and investigate possible layering effects through a parametric study. 
The reference case will be a dense suspension sheared with a uniform shear-rate
in the shear-thickening regime. The rheological quantities will be compared
to that case, with particular interest in the effective viscosity of the 
suspension. We will then study in detail the mechanisms that the inhomogeneity
triggers, with a complete analysis of the stress tensor and the breakdown of its
contribution. Finally, we will study the effect of such driving flow for the
complete flow curve. To the Authors' knowledge, this is the first three-dimensional 
complete study with this type of inhomogeneity. The novelty lies i) in the numerical 
models adopted (e.g.,~the introduction of the frictional contacts, 
described in \cref{sec:meth}), ii) in the very large number
of particles introduced for a numerical study, i.e.~$N=2^{16}$, and iii) in the full
analysis of the stress tensor of the suspension to identify key contributions to the
segregation.

\section{Methodology}\label{sec:meth}
We carried out three-dimensional numerical simulations of a dense 
suspension of rigid, spherical particles, not subject to Brownian motions, 
immersed in a Newtonian fluid with viscosity $\eta_0$.
The suspension is driven by a sum of a plain shear-flow, 
$u_1 = \overline{\dot{\gamma}} x_2$, characterised 
by the shear-rate $\overline{\dot{\gamma}}$ 
(all the quantities $\overline{\bullet}$ within the 
manuscript refer to their mean value along the shearwise direction $y$, 
while the quantities $\bullet (y)$ to their local value) 
and a sinusoidal disturbance in the velocity of the form
\begin{equation}
    u_i=A \overline{\dot{\gamma}}a_0 \sin{\left(n\kappa_0 x_j\right)},
    \label{eq:perturb}
\end{equation}
where $A$ is the amplitude of the disturbance, $a_0$ is the reference
length-scale of the problem (typically the radius of the smallest 
particle of the suspension), $\kappa_0=2\pi/L$ is the 
fundamental wavenumber of the signal (being $L$ the size of the 
computational domain), $n$ an integer that sets the wavenumber
of the signal and $x_j$ the direction of the wave. In particular, the
subscripts $i$ and $j$ span the spatial three-directions 
in a Cartesian frame of reference, where $x_1$, $x_2$ and $x_3$ 
(sometimes also referred to as $x$, $y$ and $z$) are adopted to 
identify the streamwise, shearwise and spanwise directions, and 
$u_1$, $u_2$ and $u_3$ to identify the corresponding velocity 
components ($u$, $v$ and $w$). In \cref{eq:perturb}, the constraint 
$i\neq j$ is imposed. 
\begin{figure}[t]
\includegraphics[width=0.9\textwidth]{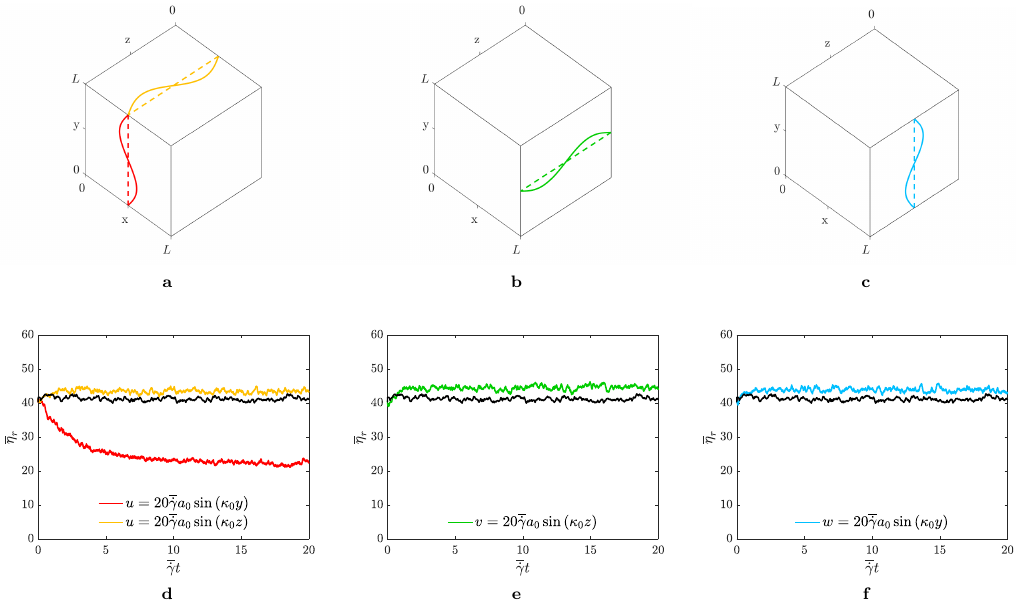}
    \caption{\label{fig:timeHis} Top row, panels \textbf{a}--\textbf{c}: sketch of 
    the sinusoidal disturbance analysed in the respective columns. 
    Bottom row, panels \textbf{d}--\textbf{f}: time-history in terms of strain $\gamma$ 
    of the relative viscosity $\overline{\eta}_r$. 
    The colours distinguish the different cases, with the black 
    solid lines referring to the values of $\overline{\eta}_r$ for the particles 
    suspended in a plain shear-flow.
    }
\end{figure}

The suspensions considered are composed of $N=2^{16}$ particles and 
are binary, i.e.\ the particles have two different sizes, with radii 
$a_1=a_0$ and $a_2=3 a_0$.
The two phases are dispersed with relative volume $V_2/V_1=0.25$.
The computational box containing the particles is a cube with
size designed to contain the desired volume fraction $\overline{\phi}=0.50$.
The uniformity of the shear-flow at the edges of the domain is 
preserved through the adoption of the 3D-periodic Lees-Edwards 
boundary conditions \citep{LEES1972}.

The numerical investigation is performed using a validated and publicly
available software, \textit{CFF-Ball-0x} \citep{MONTI2021}.
The software tackles the translational and rotational dynamics of the 
particles by solving the Newton-Euler equations,
\begin{equation}
    \begin{cases}
        m_i \dfrac{\mathrm{d}\boldsymbol{u}_i}{\mathrm{d}t} &=
    \dsum_{j=1}^{N_H} \boldsymbol{F}^H_{ij} +
    \dsum_{j=1}^{N_C} \boldsymbol{F}^C_{ij} +
    \dsum_{j=1}^{N_E} \boldsymbol{F}^E_{ij}, \\
    \amsmathbb{I}_i \dfrac{\mathrm{d}\boldsymbol{\omega}_i}{\mathrm{d}t} +
        \boldsymbol{\omega}_i \times (\amsmathbb{I}_i \boldsymbol{\omega}_i) &=
    \dsum_{j=1}^{N_H} \boldsymbol{T}^H_{ij} +
    \dsum_{j=1}^{N_C} \boldsymbol{T}^C_{ij} +
    \dsum_{j=1}^{N_E} \boldsymbol{T}^E_{ij}, \\
    \end{cases}
    \label{eq:NE}
\end{equation}
where the subscript $i$ indicates the particle $i\in [1,N]$, being
$N$ the number of particles. The right-hand side of System \eqref{eq:NE}
lists the forces and torques that are applied to the centre of mass of 
the $i$th particle, with mass $m_i$ and inertia tensor
$\amsmathbb{I}_i$ and cause a variation in the translational
and angular velocities of the particles, here denoted by the symbols
$\boldsymbol{u}_i$ and $\boldsymbol{\omega}_i$, respectively.
Such forces and torques result from the particle-fluid and particle-particle 
interactions of the $i$th particle with its $j$th neighbour. The nature 
of the interactions is indicated by the superscripts $H$, $C$ and $E$,
i.e. hydrodynamics, inelastic contacts  and electrochemical potentials,
respectively.

Here, we give a brief overview of the models used to support the three 
aforementioned contributions. For hydrodynamics, we utilized the approach 
developed by \citet{BALL1997} and \citet{MARI2014}, which considers the effects 
of a dense suspension of rigid particles in a low-Reynolds-number flow. 
These particles experience a Stokes drag and a pair-wise lubrication force 
\citep{MARI2014} that arises from the relative motion of nearby particles, 
leading to the squeezing of fluid between them. To account for these forces 
and torques, we employed a linear relationship between velocities (angular 
velocities) and forces (torques) \citep{BALL1997,MARI2014}. This relationship 
is expressed as $\boldsymbol{F}^H=-\mathbb{R}(\boldsymbol{u}-\boldsymbol{U}^\infty)$,
where $\mathbb{R}$ is a resistance matrix that takes into account 
only the dominant near-field divergent elements resulting from the 
squeeze, shear, and pump modes, as described in the work by \citet{MARI2014}. 
The far-field effects are neglected.

The stick-and-slide method \citep{LUDING2008} is used to model the contact 
contribution, which simulates inelastic contacts with spring-dashpot systems 
that become active when two spheres overlap. The force is expressed as 
$\boldsymbol{F}^C=(k_n\delta + \gamma_n\dot{\delta})\boldsymbol{n} + 
k_t\xi \boldsymbol{t}$, with the spring-dashpot systems oriented along the 
normal (centre-to-centre) $\boldsymbol{n}$ and tangential $\boldsymbol{t}$ 
directions, characterized by the spring and dashpot constants $(k_n$,$\gamma_n)$
and $(k_t$,$\gamma_t=0)$, respectively. The values $\delta$ and $\xi$ indicate 
the normal and tangential displacements of the springs, and $\dot{\delta}$ represents 
the normal projection of the overlapping velocity. The Coulomb's law, 
$\lvert \boldsymbol{F}^C_t\rvert \le \mu \lvert \boldsymbol{F}^C_n\rvert$, 
applies to the tangential contribution, where $\mu$ is the frictional coefficient, 
set to $\mu = 0.5$ for all the scenarios examined in this work, as in our 
previous study \citep{RATHEE2021} where we compared our results with experimental 
data. 

The last contribution considered in this work is the electrochemical 
properties of the suspension, modelled as a sum of an inter-particle, 
distance-decaying repulsion force and an attraction force in van der Waals 
form \citep{GALVEZ2017,SINGH2019}. The resulting force is 
$\boldsymbol{F}^E = \boldsymbol{F}^R+\boldsymbol{F}^A$, where the
where the superscripts $E,R,A$ stand for electrochemical, repulsive, 
and attractive, respectively. The repulsive contribution is given by 
$\boldsymbol{F}^R=-F_0 e^{-d/L_s}\boldsymbol{n}$, where $F_0$ is the 
magnitude of the force, $L_s$ is the screening length, and $d$ is the 
particle-particle surface distance, while the attractive force can be 
written as $\boldsymbol{F}^A = H_A\overline{a}/12(d^2+\varepsilon^2) 
\boldsymbol{n}$, where $H_A$ is the Hamaker constant, $\overline{a}$ is the harmonic 
mean radius of the two particles involved, and $\varepsilon$ is a smoothing 
term to eliminate the singularity when the two particles touch, i.e.\ $d = 0$.

The physics of the suspension described by the system of equations \eqref{eq:NE} 
is governed by several parameters. By analyzing the translational equation 
of \eqref{eq:NE} and combining the dimensional quantities, we can use 
Buckingham's $\Pi$ theorem to identify a set of non-dimensional groups that 
can be adjusted to control the desired dynamics of the suspension. 
When using the hydrodynamic scales as the fundamental independent quantities, 
we obtain three non-dimensional groups. The first is the Stokes number, 
\begin{equation}
    St = \rho_p a_0^2 \dot{\gamma}/\eta_0,
\end{equation}
which compares the time-scale of the particles with that of the hydrodynamics 
and arises from the inertial term. 
In the equation above, $\rho_p$ is the density of the particles (equal to
the density of the carrier fluid in the cases we considered), $a_0$
is the typical length scale of the system (the radius of the particles of
the smallest phase of the suspension), $\eta_0$ is the viscosity of 
the carrier fluid and $\dot{\gamma}$ is the shear-rate applied to the 
suspension. The constraint $St \ll 1$ is applied to enforce the 
inertialess regime. The second number is the non-dimensional 
The second is the non-dimensional stiffness
\begin{equation}
    \hat{k}=k_n/(\eta_0 a_0 \dot{\gamma}),
\end{equation}
which evaluates the importance of the contacts contributions relative 
to the hydrodynamic term. The constraint $\hat{k}\gg 1$ 
is imposed to ensure that the particles behave as rigid \citep{MARI2014}.
The normal spring constant $k_n$ is considered as the dominant term of 
the spring-dashpot systems, and additional constraints are applied to 
enforce this assumption. In particular, we impose
$k_t=2/7k_n$ and $\gamma_n\dot{\gamma}/k_n
\ll 1 \sim O(10^{-7})$, where the latter is the relaxation time of 
the spring-dashpot system. 
The third non-dimensional group is the equivalent shear-rate 
\begin{equation}\label{eq:hatGD}
	\hat{\dot{\gamma}} = \lvert \boldsymbol{F}^E (d=0)\rvert /(\eta_0 a_0^2 \dot{\gamma}) =
    \dot{\gamma}_0/\dot{\gamma},
\end{equation}
which measures the time-scale introduced by the electrochemical 
contribution and can be modified to impose a shear-rate dependent 
rheology on the suspension. 
A table that recaps the parameters adopted in this work is reported in the
Supplementary Material.

\begin{figure}[t]
\includegraphics[width=0.9\textwidth]{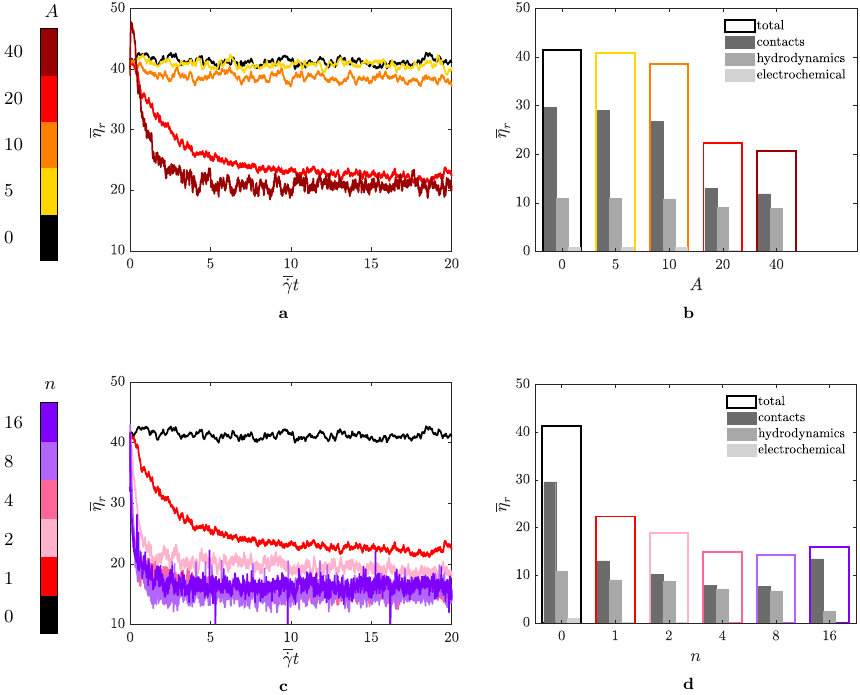}
\caption{\label{fig:etaStat} Panel \textbf{a} and \textbf{c}: time-history 
in terms of strain $\gamma$ of the relative viscosity $\eta_r$ for the 
disturbance $u=A \overline{\dot{\gamma}} a_0 \sin{\left(n \kappa_0 y \right)}$. 
Panel \textbf{b} and \textbf{d}: contributions to the mean relative viscosity
$\overline{\eta}_r$ 
accumulated at statistically-steady state of the contacts, hydrodynamics 
and electrochemical potentials (the respective colours are shown in the 
legend within the panels). The panels in the top row, i.e.\ panel \textbf{a} and 
(b), show the trends of $\overline{\eta}_r$ varying the amplitude $A$ of the disturbance
with $n=1$. The panels in the bottom row, i.e.\ panel \textbf{c} and \textbf{d}, show
the trends of $\overline{\eta}_r$ as a function of the wavenumber $n$, with the amplitude
of the sinusoidal wave fixed at $A=20$. The colorbars on the left of panel
\textbf{a} and \textbf{c} indicate the colour-scheme adopted for the amplitudes (top) and 
wavenumbers (bottom) analysed; the black solid lines refer to the reference
case of the suspension subject to a plain shear-flow. 
}
\end{figure}

\section{Results}
We start by considering a perturbation described by \cref{eq:perturb} 
with $A=20$ and $n=1$. Thus, the particles are driven by a non-uniform shear-rate, due to the
sinusoidal disturbance, but under the same mean total shear-rate.
The effects of the perturbed shear-rate on the rheology of the
suspensions are reported in \cref{fig:timeHis}.
In particular, the figure shows the relative viscosity (i.e.,\ the shear-stress ratio)
$\overline{\eta}_r=\overline{\sigma}_{12}/\eta_0\overline{\dot{\gamma}}$, where 
$\overline{\sigma}_{12}$ is the shear component of the suspension stress 
tensor $\overline{\sigma}$, as a 
function of the non-dimensional time, i.e.\ the strain 
$\gamma=\overline{\dot{\gamma}}t$, for all the possible configurations compatible
with the boundary conditions (i.e.,\ the waves $u_{2},u_{3}\sim \sin{(x_1)}$ are 
discarded since they are not consistent with the Lees-Edwards boundary 
conditions). In particular, the figure is split into
two rows, where the top one sketches the velocity disturbance and the 
bottom one shows the time-history of $\overline{\eta}_r$, keeping the same colour map
for clarity reasons. In the same figure, the black solid lines refers to
the reference case, i.e.\ the particles immersed in a plain shear-flow.
Note that, all the cases start from the same initial configuration, 
that corresponds to a bidisperse suspension with total volume 
fraction $\overline{\phi}=0.50$, and dispersion ratio $\lambda=a_2/a_1=3$. 
The suspensions are sheared with a high mean shear-rate, 
i.e.,\ $\overline{\dot{\gamma}}/\dot{\gamma}_0=1$ 
(a description of the latter quantity can be found in the Supplementary Material), 
and such configuration can be located within the shear-thickening region of the 
flow curve \citep{MARI2014,MONTI2021}.
From \cref{fig:timeHis}, we observe that three out of four cases 
do not show significant alteration in terms of $\overline{\eta}_r$, 
apart from a small increase of the relative viscosity; 
on the other hand, the remaining case manifests substantial modifications.
This case corresponds to the sinusoidal disturbance acting in the shear 
plane $x-y$, i.e.\ $u=20 \overline{\dot{\gamma}} a_0\sin{\left(\kappa_0 y\right)}$ 
(red solid line) and leads to a reduction of the effective viscosity 
by approximately $50\%$.

In the following part of this section, we will investigate the cause 
of the viscosity drop for the latter case mentioned. 
The investigation will be carried out via a parametric study aiming 
at unravelling the effect of the amplitude and of the wavenumber 
of the disturbance.

\begin{figure}[t]
\includegraphics[width=0.9\textwidth]{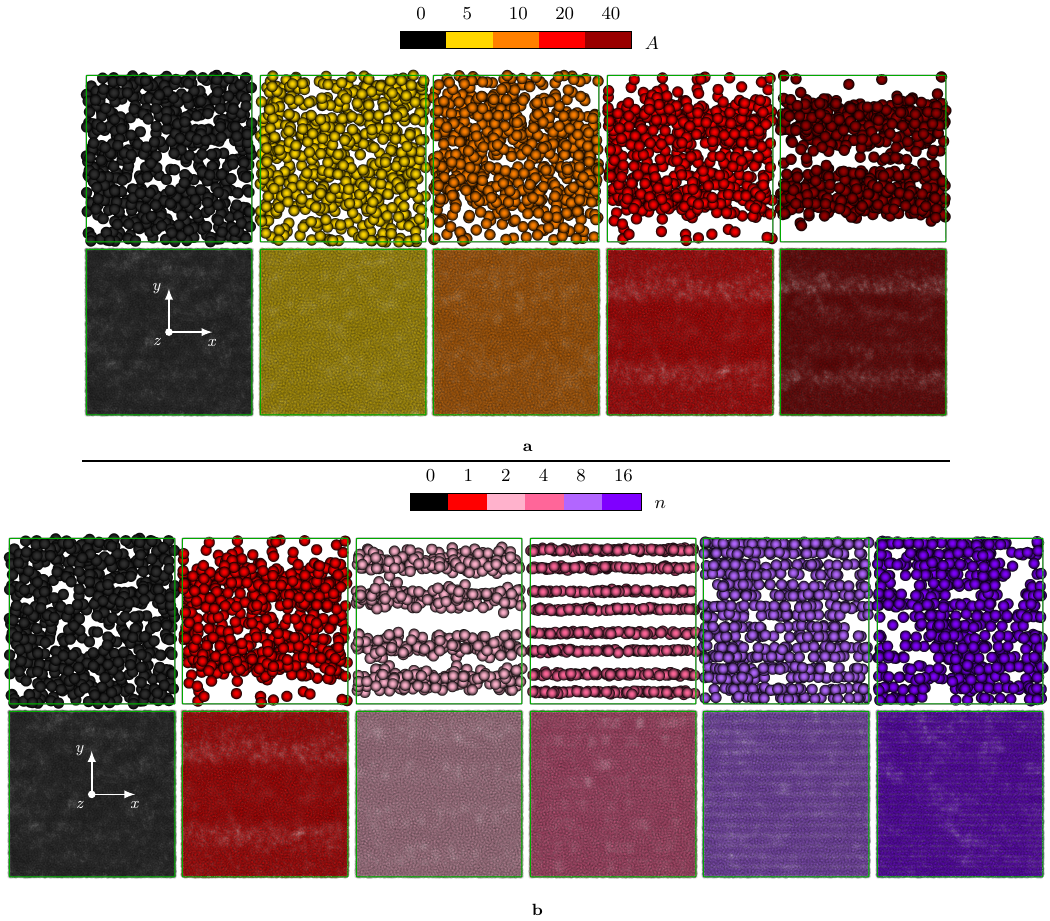}
    \caption{\label{fig:AmSnap} Panel \textbf{a}: instantaneous snapshots of the 
    suspensions at $\gamma=20$ varying the amplitude of the sinusoidal perturbation, 
    fixing $n=1$.
    Panel \textbf{b}: instantaneous snapshots of the suspensions at 
    $\gamma=20$ varying the wavenumber of the sinusoidal perturbation, 
    fixing $A=20$. 
    The top row of the subfigures shows the distribution 
    of the large particles, while the bottom row the respective distribution 
    of the small particles of the suspensions.
    The colour-scheme reflects the one adopted in the top row of
    \cref{fig:etaStat}.}
\end{figure}

\begin{figure}[t]
\includegraphics[width=0.9\textwidth]{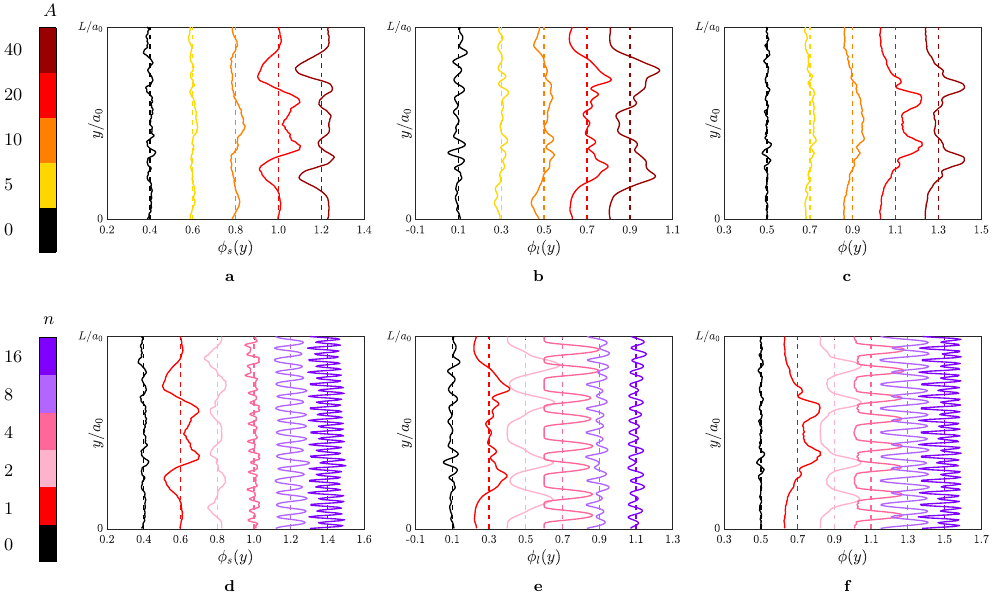}
\caption{\label{fig:localPhi} Profiles of the local volume fractions along the
shearwise direction $y$. Panel \textbf{a} and \textbf{d} show the local volume of 
the small particles $\phi_s(y)$; panel \textbf{b} and \textbf{e} show the local 
volume of the large particles $\phi_l(y)$; panel \textbf{c} and \textbf{f} show 
the total local volume fraction $\phi(y)$. The panels in the top row, i.e.\ 
panel \textbf{a}--\textbf{c}, refer to the sinusoidal
disturbance with $n=1$ and increasing amplitude $A$, as shown in the respective
legend on the left of the figure; the panels in the bottom row, i.e.\ panel 
\textbf{d}--\textbf{f}, refer to the cases with $A=20$ and increasing $n$, as shown in the 
respective legend on the left of the figure. The black profiles in each panel
refer to the reference case with no sinusoidal disturbance. Finally, the dashed
lines refer to the values $\overline{\phi}_s=0.4$, $\overline{\phi}_l=0.1$ and 
$\overline{\phi}=0.5$. Note that, the profiles for different $A$ and $n$ have been 
shifted horizontally by $0.2$ for visual purpose.
}
\end{figure}

\Cref{fig:etaStat} reports the trends of the relative viscosity
obtained varying the amplitude $A$ (top row) and the wavenumber
$\kappa=n \kappa_0$ (bottom row) of the disturbance.
At first, we analyse the effect of the amplitude $A$; in particular, 
in the top row of \cref{fig:etaStat}, 
we show the time-history 
of the relative viscosity $\overline{\eta}_r$ (left panel), and we report the mean 
values of $\overline{\eta}_r$ collected at statistically steady-state, together with 
their contributions breakdown (right panel).
Starting from the zero-amplitude case, i.e. uniform shear flow (black line), 
we observe that the effective viscosity monotonically decreases 
by increasing the amplitude, eventually saturating for $A>20$.
The reduction rate is strongly non-linear, with very little changes of 
$\overline{\eta}_r$ for $A \le 10$, followed by a sudden decrease. This is clearly
visible from the mean values of $\overline{\eta}_r$ at statistically steady-state 
(right panel, total value in the legend).
Therefore, as a first result, the present behaviour suggests the existence of 
a threshold value which activates the mechanism of the reduction of the relative 
viscosity. Note that, the final value of the effective viscosity is reached after a long 
transient, i.e.\ $\gamma \sim 15$, and that increasing the amplitude of the disturbance
contributes to shorten the convergence 
time of the rheological response of the suspension.


Next, we report the trends of the relative viscosity
obtained varying the wavenumber $\kappa=n \kappa_0$ of the disturbance.
As done for the amplitude $A$, in the bottom row of \cref{fig:etaStat} 
we show the time-history 
of the relative viscosity $\overline{\eta}_r$ (left panel), and we report
the mean values of $\overline{\eta}_r$ collected at statistically steady-state, 
together with their relative contributions (right panel).
Similarly to the previous cases, starting from the uniform shear flow
(black line), we observe that the effective viscosity non-linearly 
decreases when a disturbance with higher harmonics is applied to the 
suspension, reaching a minimum value between $n=4$ and $n=8$. 
For higher harmonics (e.g.,\ $n=16$), the relative viscosity starts
increasing again, thus suggesting the presence of an optimum
harmonic that minimizes the value of the relative viscosity of the 
suspension. This tendency is also confirmed by the mean values 
of the relative viscosity at statistically steady-state. 
Another outcome suggested by the time-history of the relative viscosity 
is that the harmonics with higher wavenumbers significantly reduce 
the convergence time (as done by the higher amplitudes), nevertheless increasing
the overall level of fluctuations,
and showing the presence of an optimum value (i.e.,\ faster convergence rate) 
for wavenumbers included between $n=4$ and $n=8$.

To understand the cause of the viscosity reduction, we give a closer look to the 
several stress contributions to the effective viscosity. 
In panel \textbf{b} and \textbf{d} of \cref{fig:etaStat}, we compare the reference case
(plain shear) to the suspensions driven by the non-uniform shear flow varying 
the amplitude $A$ of the sinusoidal disturbance (panel \textbf{b}, with $n=1$) and its 
wavenumber $n$ (panel \textbf{d}, with $A=20$).
The three components of the stress $\overline{\eta}_r$ are shown: 
the hydrodynamic contribution which includes the Stokes drag, the lubrication and 
the Newtonian component of the carrier fluid; the interactions caused by the 
electrochemical potentials and the stresses due to the particle-particle contacts. 
The suspensions analysed in the parametric study are located within the
shear-thickening region of the flow curve, i.e.,\ $\overline{\dot{\gamma}}/\dot{\gamma}_0=1$,
thus, the contributions to the shear-stress related to the contacts and the 
hydrodynamics are dominant (with the former being more important), while the 
electrochemical potentials play no effective role. Within the parameters chosen 
in this work, we observe that the reduction of $\overline{\eta}_r$ (for the cases with 
$A\ge 20$ and $n\ge 1$) is mostly due to a strong decrease of the contacts 
contribution to the shear-stress, while the hydrodynamics remains almost 
unchanged. Worth noticing is the case with $A=20$ and $n=16$, where also the hydrodynamics 
contribution is significantly reduced compared to the other cases analysed.
\begin{figure}[t]
%
\includegraphics[width=0.9\textwidth]{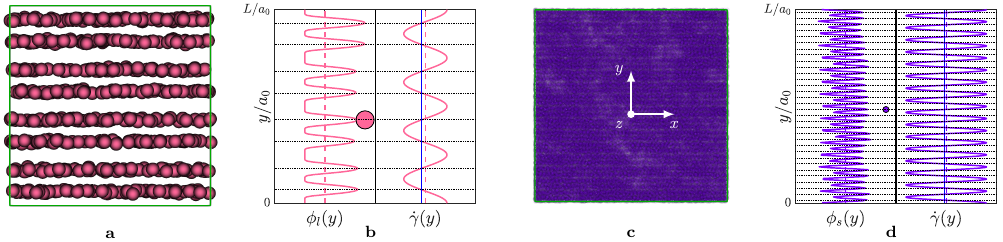}
    \caption{
    \label{fig:accumulPoints} Panel \textbf{a} and \textbf{c}: instantaneous snapshot
    of one of the two dispersed phases of the suspension; panel \textbf{b} and \textbf{d}:
    plots of the local volume fraction along the shearwise direction $y$ 
    of the respective phase, alongside the 
    local profile of the shear-rate applied to the suspension (right side).
    Panel \textbf{a} and \textbf{b}: case $A=20$ and $n=4$, with the snapshot of the 
    distribution of the large phase in the suspension; panel \textbf{c} and \textbf{d}:
    case $A=20$ and $n=16$, with the snapshot of the distribution of the 
    small phase in the suspension. 
    The two plots in panel \textbf{b} and \textbf{d} are split by the vertical, black solid 
    line in the middle of the figure; the other vertical straight lines 
    represent, from left to right: the mean value of the volume fraction of the 
    particles ($\overline{\phi}_l=0.1$ or $\overline{\phi}_s=0.4$, pink/violet dashed line); 
    $\dot{\gamma}(y)=0$ (blue solid line) and the mean value of the total shear-rate 
    applied to the suspension (pink/violet dashed line). 
    The dotted horizontal lines outline the $y$ location of
    $\dot{\gamma}(y)=0$. Finally, the pink/violet circle in 
    the graphs sketches a large/small particle of the suspension.
	}
\end{figure}

\begin{figure}[t]
\includegraphics[width=0.9\textwidth]{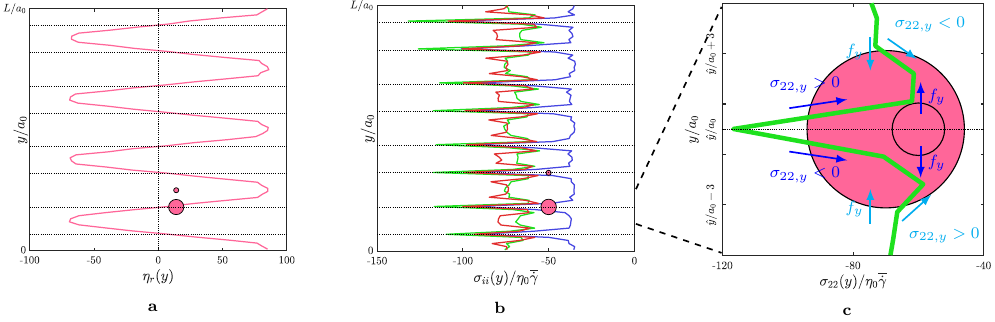}
    \caption{\label{fig:localStr} Panel \textbf{a}: local distribution of the
    mean shear-stress ratio $\eta_r(y)$ along the shearwise direction $y$.
    The vertical dotted line highlights $\eta_r(y)=0$; the two circles
	sketch the two types of particles dispersed in the suspension.
    Panel \textbf{b}: local distribution of the mean normal stresses, $\sigma_{11}(y)$ 
    (red line), $\sigma_{22}(y)$ (green line) and $\sigma_{33}(y)$ (blue line),
    along the shearwise direction $y$. Panel \textbf{c}: zoom of $\sigma_{22}(y)$ 
	in one of the regions of panel \textbf{b} bounded by the dashed lines.
    The location $\hat{y}/a_0$ corresponds to the zero-shear location $\dot{\gamma}(y)=0$.
    The two circles represent the particles of the two phases of the suspension. 
    The arrows show the directions of the gradients of the 
    normal stress $\sigma_{22,y}$ and the respective forces they generate on
    the particles. In all the panels of the figure, the horizontal 
    dotted lines indicate the regions at $\dot{\gamma}(y)=0$. The case 
    considered has a sinusoidal perturbation with $A=20$ and $n=4$.
	}
\end{figure}

To highlight the origin of the relative viscosity reduction, we show in 
\cref{fig:AmSnap}, panels \textbf{a} and \textbf{b}, the instantaneous 
snapshots of the suspensions studied when a statistically steady-state
is reached.
In particular, the two panels are 
organised in a $5\times 2$ (or $6\times 2$) matrix as follows:
from left to right, the amplitude (or the wavenumber) increases, from 
$A=0$ ($n=0$) to $A=40$ ($n=16$); in the top row, the distribution of 
the larger dispersed phase is shown while, in the bottom row, the location 
of the smaller dispersed phase is displayed.
The two panels qualitatively show the effect of the sinusoidal disturbance on the
suspension: the large particles tend to separate from the small ones, forming 
parallel layers normal to the shearwise (i.e.\ $y$) direction. 
The number of layers of large particles seem to be correlated with the 
wavenumber of the disturbance, while the role of the amplitude appears to be related 
to a threshold that activates the demixing of the phases,
as previously postulated from the mean values of the relative viscosity 
(i.e.,\ \cref{fig:etaStat}).

Quantitatively, this becomes clear when showing the
small and large particles concentration along the $y$-coordinate, as 
pictured in \cref{fig:localPhi}
{ (note that the symmetry of the profiles is discussed in the Supplementary
 Material)}. The figure shows, within each row, the volume fraction of 
the small particles $\phi_s(y)$, that of the large particles $\phi_l(y)$ and 
the total volume fraction $\phi(y)$, as a function of the amplitude of the 
disturbance with wavenumber $n=1$ (top row), and as a function of the 
wavenumber $n$, fixing the amplitude to $A=20$ (bottom row). Note that
the local volume fractions oscillate around the values $\overline{\phi}_s = 0.4$, 
$\overline{\phi}_l=0.1$ and $\overline{\phi}=0.5$, highlighted by the dashed lines; 
therefore, the labels on the abscissas simply indicate the aforementioned values 
shifted by an offset to show all the curves on a single graph.
From the top row of \cref{fig:localPhi}, we observe the progressive formation 
of two regions with high large particles concentration $\overline{\phi}_l$, with 
the peaks expanding monotonically by increasing the amplitude of the disturbance. 
On the other hand, in the bottom row of \cref{fig:localPhi},
as $n$ grows we observe an increase of the number of layers where the large 
particles are collected, thus suggesting that the number of accumulation 
regions is controlled by the wavelength of the disturbance with the amplitude 
controlling only the amount of the relative concentration. This 
quantitative analysis corroborates what observed in the snapshots shown
in \cref{fig:AmSnap}.
Panel \textbf{d} and \textbf{e} of \cref{fig:localPhi} also show that, for small
wavenumbers, i.e.,\ $n\le 8$, the large particles accumulate into well-separated 
layers while, for large wavenumbers, i.e.,\ $n>8$, the small ones do. 
This is a consequence of the size of the particles compared to the wavelength 
of the perturbation: in the former case, 
the large particles have a diameter comparable to or smaller than the 
wavelength of the sinusoidal wave, defined as $\Lambda=1/(n\kappa_0)$,
while in the latter case, the small ones do. Indeed, when
the wavelength of the disturbance is much smaller than the particle size, 
its effect is filtered out by the 
particle size, with the particles feeling the disturbance as noise in the 
limit of $\Lambda \ll a_1,a_2$. 
This becomes clear when looking at the local distributions of the 
large particles for $n=8$ and $n=16$ in panel \textbf{e} of \cref{fig:localPhi}: 
the two curves show the same number of peaks ($16$),
meaning that the effect of the wavelength on the large particles had reached
a maximum and saturated. An estimate of the maximum 
value of the wavenumber felt by particles of radius $a_i$ (with $i=1,2$) can be
easily computed as
\begin{equation}
    n_{max} \approx \dfrac{1}{2}\dfrac{L}{2 a_i}, 
\label{eq:n_wnMax}
\end{equation}
which for the large particles we consider gives $n_{max} \approx 7$, comparable to $n=8$.
The wavenumber $n_{max}$ is half the ratio between $L$ and $d$ because of 
the points of accumulation of the particles, 
which are located in the proximity of the zero-shear values, as it can be 
seen in \cref{fig:accumulPoints}. In particular, panel \textbf{a} 
of the figure shows a snapshot at statistically steady-state of the large 
particles for the case $A=20$ and $n=4$, while in panel \textbf{b} the mean local 
volume fraction of the large particles (left side) together with the shape 
of the local shear-rate (right side) are plotted.
From panel \textbf{a} and \textbf{b}, we can see how the large particles accumulate 
in parallel layers around the locations of the zero-shear regions 
(dotted horizontal lines in panel \textbf{b}). On the other hand, 
when considering disturbances with high wavenumbers, 
the smaller particles are reorganizing in parallel layers, as 
can be observed in panel \textbf{c} and \textbf{d} of \cref{fig:accumulPoints}.
The two panels mirror panel \textbf{a} and \textbf{b} for the case $A=20$ and $n=16$.
For this case, it is clear that the small particles collect around 
the zero-shear regions.

\begin{figure}[t]
\includegraphics[width=0.9\textwidth]{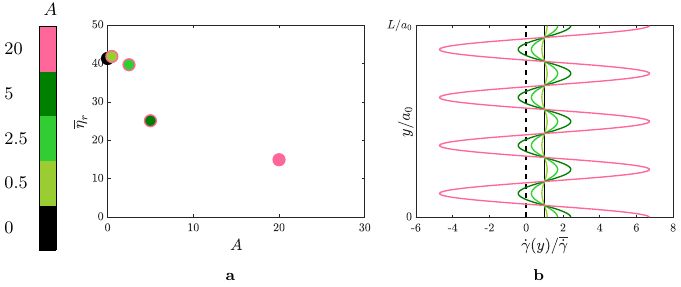}
\caption{\label{fig:amN4} Wavenumber $n=4$.
    Panel \textbf{a}: mean relative viscosity $\eta_r$ at
    statistically steady-state for the disturbance 
    $u=A \overline{\dot{\gamma}} a_0 \sin{\left(4 \kappa_0 y \right)}$, varying
    the amplitude $A$ of the disturbance. Panel \textbf{b}: profiles of the 
    shear-rate applied to the suspensions along the shearwise direction
    $y$. The colour-scheme indicates the amplitude imposed and reflects 
    the one used in the colorbar. The black dashed line is the 
    zero shear-rate.
    }
\end{figure}
\begin{figure}[t]
\includegraphics[width=0.9\textwidth]{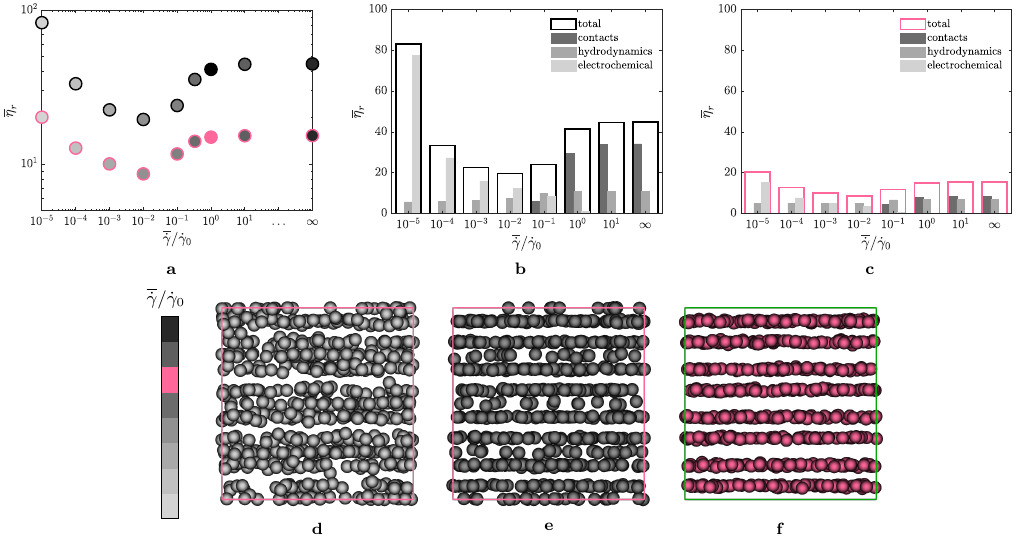}
    \caption{\label{fig:gd} Top row, panels \textbf{a}--\textbf{c}. 
       Panel \textbf{a}: mean relative viscosity $\eta_r$ at
       statistically steady-state for suspensions driven by a plain shear-flow
       (markers with a black edge) and for the suspension subject to the disturbance 
       $u=20 \overline{\dot{\gamma}} a_0 \sin{\left(4 \kappa_0 y \right)}$ (markers with a
       pink edge) as a function of the shear-rate $\overline{\dot{\gamma}}/\dot{\gamma}_0$.
       Panel \textbf{b} and \textbf{c}: contributions to the mean relative 
       viscosity $\overline{\eta}_r$  
       accumulated at statistically-steady state of the contacts, hydrodynamics 
       and electrochemical potentials (the respective colours are shown in the 
       legend within the panels), as a function of the shear-rate 
       $\overline{\dot{\gamma}}/\dot{\gamma}_0$, for the suspension driven by a
       plain shear-flow (panel \textbf{b}) and for the suspension subject to the 
       sinusoidal disturbance (panel \textbf{c}).
       Bottom row, panels \textbf{d}--\textbf{f}: 
       instantaneous snapshots of the large particles
       of the suspensions at statistically-steady state varying 
       $\overline{\dot{\gamma}}/\dot{\gamma}_0$.
       Panel \textbf{d}: $\overline{\dot{\gamma}}/\dot{\gamma}_0=10^{-5}$; 
       panel \textbf{e}: $\overline{\dot{\gamma}}/\dot{\gamma}_0=10^{-2}$; 
       panel \textbf{f}: $\overline{\dot{\gamma}}/\dot{\gamma}_0=10^{0}$.
       The colours in panel \textbf{a} and panels \textbf{d}--\textbf{f} refer
       to the colormap in the figure. Note that the colour pink can be changed to
       black in panel \textbf{a} to show the non-perturbed case.
       The suspensions reported are subject to the sinusoidal perturbation 
       with $A=20$ and $n=4$.
       }
\end{figure}
Next, to understand the mechanism of the particles accumulation, 
in \cref{fig:localStr} we show the relevant components of the local relative 
mean stress tensor for the case $A=20$ and $n=4$ at statistically steady-state.
In particular, panel \textbf{a} and \textbf{b} show the local relative viscosity
(i.e.,\ the local shear-stress ratio)
and the local relative normal components of the stress tensor, respectively,
along the shearwise direction $y$. Note that the integral value of the local
shear-stress ratio along the shear-wise direction $y$ corresponds to the 
mean relative viscosity. In these panels, the dotted horizontal 
lines outline the zeros of the total shear-rate 
profile, i.e.,\ $\dot{\gamma}(y)=0$. 
The profile of the local shear-stress ratio promptly highlights
that the suspension is sheared in two opposite directions (opposite signs of 
$\eta_r(y)$) following the shape of the imposed shear-rate, with 
regions of zero shear-stress in the proximity of the large particles collection 
points (i.e.,\ $\dot{\gamma}(y)=0$). 
Focusing on the normal stresses, instead, it is clear that, in the regions
where the large particles accumulate, there are peaks of high-compressive 
stresses along the three directions.
For $\sigma_{11}(y)$ (red line) and $\sigma_{22}(y)$ (green line)
those peaks are enclosed by two adjacent local maxima,
whose distance is comparable with the diameter
of the large particles. 
A zoom on one of the regions enclosing the three aforementioned
peaks (one minimum and two maxima) is
shown in panel \textbf{c} of \cref{fig:localStr}. In particular, only the normal 
stress along the shearwise direction $\sigma_{22}(y)$ is reported.
Furthermore,  in the figure we also sketch the force $f_y$ that the gradients of the normal stress
$\sigma_{22,y}$ generate on the particles: we notice how the gradients enclosing
the minimum peak tend to push away the particles from the zero-shear regions, 
while the two local maxima have an opposite effect.
From the sketch, it is clear that small particles are effectively pushed out
of the zero-shear regions, while the large particles are trapped
by the two small stabilising peaks of $\sigma_{22}(y)$. Indeed,  the
effect of the stronger peak in the proximity of
the zero-shear regions is filtered out by the large particles due to its size.
This explains the mechanism of the particles migration for this 
configuration: the small particles migrate away from the zero-shear regions 
and, in turn, push the large particles in the accumulation points, eventually
locking them there by acting with compressive forces on the extrema of the
large particles.

One of the reasons that trigger the phase separation and the demixing process
can be sought in the strength of the forcing term that imposes positive and 
negative shear-rates to the suspension. To investigate this, we carry out
a further parametric study where we choose the case with the streamwise wave
having $n=4$ (that showed an efficient demixing) and we vary the amplitude
of the wavy disturbance.
Panel \textbf{a} of \cref{fig:amN4} reports the values of the relative 
viscosity collected at statistically steady-state for the cases with 
amplitude $A=[0.5,2.5,5,20]$, compared to the reference case, i.e.\ the suspension
subject to a plain shear flow (black marker with $A=0$). The figure shows a similar
non-linear trend seen for the cases with $n=1$, with a strong reduction of the
relative viscosity between $A=2.5$ and $A=5$. In panel \textbf{b}, instead, we
report the profiles of the applied shear-rates along $y$. 
Note that the case with $A=5$ is the first one to cross the zero-shear 
line (dashed black line). This suggests that, when the disturbance is high 
enough to produce negative shear-rates, the demixing is activated 
causing a sudden reduction of $\overline{\eta}_r$. Note that, being the total shear-rate
$\dot{\gamma}(y)=\overline{\dot{\gamma}} + A \overline{\dot{\gamma}} a_0 n \kappa_0 
\cos{(n \kappa_0 y)}$,
either a disturbance with a strong amplitude or one with a large wavenumber 
(or a combination of the two) may cause the demixing of the dispersed phases.

%
The final question of this work concerns the extension of the demixing process
seen within the manuscript for the other rheological regimes of the dense suspensions, 
i.e.\ for any point belonging to the whole flow-curve ($\overline{\eta}_r$ 
as a function of the shear-rate). To introduce the shear-rate dependency,
we tune the amplitude of the electrochemical as in \cref{eq:hatGD}
\citep[see also][]{MARI2014}. To determine
the Hamaker constant and the intensity of the repulsive forces separately, we
set the ratio $\lvert \boldsymbol{F}_R \rvert / \lvert \boldsymbol{F}_A \rvert = 9$
when two particles are at contact $d_{ij}=0$.
Panel \textbf{a} of \cref{fig:gd} shows the values of the relative viscosity
collected at statistically steady-state for the suspensions immersed in a plain shear 
flow (markers with the black edge) and for the suspensions subject to the sinusoidal
disturbance with $A=20$ and $n=4$ (markers with the pink edge), varying the shear-rate 
$\overline{\dot{\gamma}}/\dot{\gamma}_0$.
Panel \textbf{b} and \textbf{c}, instead, show the decomposition
of the relative viscosity in the three main contributions (as seen in 
\cref{fig:etaStat} already) for the suspension subject to the plain shear-flow and 
the sinusoidal shear-flow, respectively. As it can be seen, the 
relative viscosity is always lower when a sinusoidal shear is imposed to the 
suspension for any $\overline{\dot{\gamma}}/\dot{\gamma}_0$, meaning that the demixing is 
always activated, despite the different nature of the contributions that govern
the rheology of the suspensions in the various regimes, as it can be seen from 
the histograms. In particular,  while for high $\overline{\dot{\gamma}}/\dot{\gamma}_0$
the reduction of the relative viscosity is due to a reduction of the contacts, 
as discussed before, for low $\overline{\dot{\gamma}}/\dot{\gamma}_0$ this is due 
to a reduction of the electrochemical interactions.
This suggests that the segregation phenomenon in this case is driven by the 
response of the suspension to the high shear introduced by the sinusoidal wave, 
that in turns activates the selective process carried out by the normal stresses.
The effect of the different contributions controlling the particle dynamics
in the various regimes, however, can be spotted from the instantaneous
snapshots of the large particles at statistically-steady state, as shown in the bottom row,
i.e.\ panels \textbf{d}--\textbf{f}, of \cref{fig:gd}. 
In particular, panel \textbf{d}--\textbf{f} show the 
instantaneous distributions
of the large particles for $\overline{\dot{\gamma}}/\dot{\gamma}_0=10^{-5}$, 
$\overline{\dot{\gamma}}/\dot{\gamma}_0=10^{-2}$ and 
$\overline{\dot{\gamma}}/\dot{\gamma}_0=10^{0}$, respectively. 
These three particular scenarios have been selected based on panel \textbf{c} of \cref{fig:gd} 
to show the effect of the different contributions: in fact, in the configuration of panel 
\textbf{d}, the dominant contribution is given by the electrochemical potentials (i.e., \ forces 
interactions along the centre-to-centre direction); in that of panel \textbf{e}, the dominant
contribution is given by the hydrodynamics, while in the condition of panel \textbf{f} by the 
normal and frictional contacts.
From the figure, it can be seen that when the contributions to the stress tensor  
are dominated by forces along the centre-to-centre direction (such as in panel \textbf{d}), the 
large particles accumulate in the region with positive shear-rate (see the pink profile in
panel \textbf{b} of \cref{fig:amN4}) while, when forces with tangential direction start appearing
(gradually increasing from panel \textbf{e} to panel \textbf{f}), the large particles tend 
to stabilise around the zero-shear regions only.

\section{Discussion}
We have performed numerical simulations of dense binary suspensions with a large 
dispersion ratio, under a non-uniform shear flow,
consisting of the combination of a linear shear and a sinusoidal disturbance. First, we find that
the only disturbance altering the rheology of the suspension is the one acting on the 
streamwise component of the velocity in the shear-plane.  By varying the amplitude and 
wavenumber of the perturbation, we discover that the demixing process can be triggered when 
the local shear-rate becomes negative. When the phases separate, the relative viscosity
reduces in the whole flow curve.
We explain the process by analysing the full stress tensor of the suspension:
we find that large particles are locked in the accumulation points by compressive forces,
while the small ones are pushed out by the gradients of the normal stresses. 
With this analysis, we highlight the importance of the full
stress tensor to unravel the complex rhelogical behaviour of dense suspensions.

The phenomenon of particle separation was examined throughout the entire rheological 
curve under optimal conditions ($A=20$ and $n=4$). Regardless of the shear rate 
applied to the suspension, the suspension undergoes phase-separation, although the 
relative significance of the factors contributing to shear stress have a different
nature. On the right side of the flow curve, contact interactions dominate, 
while in the shear-thinning region, electrochemical potentials take precedence. 
This demixing effect leads to the accumulation of larger particles 
(for $n \le 4$ in our study) in different regions along the direction of shear:
the zeros of the shear-rate for conctact-dominated suspesions, positive shear-rate
for electochemical potentials-dominated interactions.

It is important to underline that, to properly understand the demixing,
considering the finite size of the particles is essential to filter out and
select the proper information from
the gradients of the normal stress tensor. Therefore, caution should be used when
employing point-size particles models to explain or analyse such phenomena.

The outcomes of our analysis have far-reaching implications within several 
scientific disciplines, specifically rheology, microfluidics, biofluidics, 
and granular flows. By delving into these areas, our research contributes 
to a deeper understanding of the behaviour of various materials and fluids 
under different conditions.

In the field of rheology, our findings provide a foundational basis for 
developing a comprehensive rheological model that accurately captures 
the complexities of real-world situations. Traditionally, rheological 
models have overlooked structural inhomogeneity or natural disturbances, 
which can have substantial effects on material behaviour. 
However, our study addresses this limitation by considering these factors, 
thus offering a more realistic depiction of rheological phenomena. 
Moreover, our research extends its impact to the realm of microfluidics, 
where the manipulation and control of fluids at the microscale are essential. 
By incorporating our insights into microfluidic systems, researchers and 
engineers can gain a better understanding of how spatial disturbances 
influence the fluid flow and behaviour, enabling them to design more 
efficient and precise microfluidic devices. Additionally, our research 
contributes to the study of granular flows. By considering spatial 
disturbances and their effects on granular materials, we offer a more 
comprehensive framework for modelling and analysing granular flow 
behaviour, thereby facilitating improved process design and optimisation.

An exciting future direction for our study involves exploring the 
coupling of spatial disturbances with the more commonly studied 
temporal disturbances. By incorporating both spatial and temporal 
factors into our model, we can take a significant stride towards 
simulating and understanding real-world conditions more accurately. 
This advancement would enable researchers to investigate how 
disturbances propagate and interact over time, providing a more 
holistic understanding of complex systems and further bridging 
the gap between theoretical models and real-world scenarios.

\section*{Acknowledgments}
All authors gratefully acknowledge the support of Okinawa Institute of Science and Technology Graduate University (OIST) with subsidy funding from the Cabinet Office, Government of Japan. The authors also acknowledge the computer time provided by the Scientific Computing section of the Research Support Division at OIST.

\section*{Author contributions}
M.E.R. conceived the original idea and planned the research. A.M. developed the code and performed the numerical simulations. All authors analyzed the data and outlined the manuscript content. A.M. wrote the manuscript with feedback from all authors.

\section*{Code and data availability}
The code used for the present research is a fully validated
software, \textit{CFF-Ball-0x}, publicly available at 
\url{https://github.com/marco-rosti/CFF-Ball-0x}. All the reported results
can be reproduced using this code and the information provided in the text.
All data are available from the authors upon reasonable request.

\section*{Competing interests}
The authors declare no competing interests.

\end{document}


\preprint{APS/123-QED}

\title{Supplementary information to: Dense bidisperse suspensions under non-homogeneous shear}

\author{Alessandro Monti$^*$ (alessandro.monti@oist.jp)}
\author{Marco Edoardo Rosti$^*$ (marco.rosti@oist.jp)}
\affiliation{Complex Fluids and Flows Unit, Okinawa Institute of Science and Technology Graduate University, 1919-1 Tancha, Onna-son, Okinawa 904-0495, Japan}%

\date{\today}
\maketitle

\section*{Supplementary material to: Methods}
In this appendix, we report the expressions of the forces acting on the particles considered in this study,
and provide more details on the numerical method.
In particular, in our model we consider the hydrodynamics, inelastic contacts  and electrochemical potentials,
and indicate them by the superscripts $H$, $C$ and $E$,respectively. 

Concerning the hydrodynamics, dense suspensions of rigid particles immersed 
in a low-Reynolds-number flow are subject to a Stokes drag and a pair-wise, 
short-range lubrication force \citep{MARI2014} caused by the reaction of
the fluid squeezed in the narrow gaps between two neighbouring particles.
This contribution is modelled in our software as a linear 
relationship between the forces and the velocities,
$\boldsymbol{F}^H=-\amsmathbb{R}(\boldsymbol{u}-\boldsymbol{U}^\infty)$,
where $\amsmathbb{R}$ is a resistance matrix that is obtained by
neglecting the far-field effects and considering only the dominant
near-field (threshold gap between particles $\delta_{lub} < a_0$) divergent 
elements coming from the squeeze, shear and pump
modes \citep{BALL1997,MARI2014}. More details on the coefficients of the 
resistance matrix adopted can be found in \citet{MONTI2021}.

The contacts, instead, are replicated with the stick-and-slide method
\citep{LUDING2008}, that mimics inelastic, frictional contacts between a 
pair of slightly overlapping particles with spring-dashpot systems placed in 
the normal (centre-to-centre) and tangential directions.
Considering two particles, $i$ and $j$, the stick-and-slide model
can be written as
\begin{subequations}
\begin{align}
    \label{eq:contact0}
    &\boldsymbol{F}^{C}_{n,ij} = k_n \delta_{ij} \boldsymbol{n}
                              + \gamma_n \boldsymbol{u}_{n,ij},\\
    &\boldsymbol{F}^{C}_{t,ij} = k_t \xi_{ij}\boldsymbol{t}, \\
    &\boldsymbol{T}^{C}_{ij} = a_i \boldsymbol{n} \times
                            \boldsymbol{F}^{C}_{t,ij},
\end{align}
\end{subequations}
where the vectors $\boldsymbol{n}$ and $\boldsymbol{t}$, together 
with the subscripts $n$ and $t$, indicate the directions of the 
force (i.e.\ normal and tangential) and the parameters $k_n$, 
$k_t$ and $\gamma_n$ are the spring constants and the normal 
damping constant, respectively.
The velocity $\boldsymbol{u}_{n,ij}$ is the projection of the
relative velocity vector between the particles $i$th and $j$th 
projected in the normal direction, while $\delta_{ij}$ and $\xi_{ij}$ 
represents the displacement of the normal and tangential spring. The 
latter is an integral length that considers the history of the contact 
\citep{LUDING2008} and its value is such that the tangential force 
satisfies the Coulomb's law 
$|\boldsymbol{F}^{C}_{t,ij}| \leq \mu_C |\boldsymbol{F}^{C}_{n,ij}|$, 
where $\mu_C$ is the friction coefficient (in this work, 
$\mu_C=0.5$ for all the suspensions simulated).

Finally, the contribution from the electrochemical potentials is modelled 
as a combination of a distance-decaying repulsive force (with the distance-decay 
length scale set by the screening length) and an attractive force expressed 
in the van der Waals form \citep{GALVEZ2017,SINGH2019},
\begin{subequations}
\begin{align}
    &\boldsymbol{F}^R_{ij} =
        - F_0 e^{-d_{ij}/L_S}\,\boldsymbol{n},
        & \text{if $d_{ij}\geq 0$}, \label{eq:rep}\\[15pt]
    &\boldsymbol{F}^A_{ij} =
        \dfrac{H_A \overline{a}}{12(d_{ij}^2 + \varepsilon^2)}
         \,\boldsymbol{n} & \text{if $d_{ij}\geq 0$}. \label{eq:att}
\end{align}
\end{subequations}
In \cref{eq:rep}, $F_0$ is the amplitude of the repulsive force, 
${\overline{a}} = 2 a_i a_j/(a_i+a_j)$ the harmonic radius and $L_S$ 
is the screening length that controls the decay of the force. 
In \cref{eq:att}, instead, $H_A$ is the
Hamaker constant and $\varepsilon$ is a small regularisation term
(generally $\epsilon \approx 0.01\overline{a}$) introduced to avoid the
singularity at contact, when the surface distance between the
particles $d_{ij}=0$.

All these contributions induce mechanical stress in the suspension
and the stress tensor $\sigma$ can be evaluated through the 
stresslet theory \citep{GUAZZELLI2011}. From $\sigma$, all the
rheological quantities (relative viscosity $\eta_r$, first and second 
normal stress differences $N_1$ and $N_2$ and pressure $\Pi$) can be 
extrapolated.

The physics of the suspension described by the Newton-Euler
equations is dominated by three non-dimensional
groups obtained by applying the Buckingham Pi theorem and choosing
as fundamental quantities the properties that describe the hydrodynamic
force, i.e. $F^H\propto \eta_0 a_0^2 \overline{\dot{\gamma}}$. 
The non-dimensional groups are listed in the main manuscript.\\
In equation (4) of the manuscript, $k_n$ approximates the equivalent
stiffness of the spring-dashpot systems,
\begin{equation}
    k_{eq} = k_n \left( 1 + \dfrac{k_t}{k_n} + 
             \dfrac{\gamma_n \overline{\dot{\gamma}}}{k_n} \right),
    \label{eq:keq}
\end{equation}
with the additional constraints
$k_t \ll k_n$ and $\gamma_n\overline{\dot{\gamma}}/k_n\ll 1$ \citep{MARI2014}.
In equation (5) of the manuscript,
we defined a new shear-rate, $\dot{\gamma}_0=F^E/\eta_0 a_0^2$,
being $F^E =  |\boldsymbol{F}^E (d_{ij}=0)| = F^R + F^A$ 
(with $F^R = |\boldsymbol{F}^R_{ij} (d_{ij}=0)|$ and
$F^A = |\boldsymbol{F}^A_{ij} (d_{ij}=0)|$). in $F^E$, the 
dominant contribution is given by the 
repulsive forces, with the choice $F^R = 0.9 F^E$ and
$F^A = 0.1 F^E$.
This non-dimensional group represents the additional time-scale introduced 
by the electrochemical contribution and it can be thought of as the equivalent 
of the P{\'e}clet number for the Brownian-suspensions; thus, tuning the
module of the repulsive force $F^R$ and the Hamaker constant 
appropriately, a shear-rate dependency can be imposed to the suspension.
Note that in the manuscript, at times we omitted the subscript $ij$ for the sake
of readability.

From a computational viewpoint, the governing equations (2) in the manuscript,
are discretized and advanced in time with the modified velocity-Verlet 
explicit scheme \citep{GROOT1997}.
The modified velocity-Verlet scheme is second-order accurate in time and, 
being explicit, it requires a time-step $\Delta t$ able to capture the 
shortest dynamics of the systems, typically established by the stiffness of the
contacts; a good indicator of the time-step required can be obtained by 
building the time constant from the normal spring-dashpot system, i.e.\
$\Delta t\overline{\dot{\gamma}}=\gamma_n\overline{\dot{\gamma}}/k_n$. Finally,
as already mentioned above, the computational domain is a cubic box of 
size $L^3$. The shear-rate $\overline{\dot{\gamma}}$ is applied along the $y$ 
direction, with the Lees-Edwards boundary conditions \citep{LEES1972}
preserving the ideality of the flow, removing the effect of the walls.

The set of parameters for carrying out these simulations are listed in
\cref{tab:sr}. Note that, for the sake of simplicity and completeness, 
$a_0 = a_1 = 1\,\,\mathcal{L}$, $\overline{\dot{\gamma}} = 1\,\,\mathcal{T}^{-1}$ and 
$\rho_p = 1\,\,\mathcal{M}\,\mathcal{L}^{-3}$. Here
$\mathcal{L}$, $\mathcal{T}$ and $\mathcal{M}$ are the typical scales for length, time and
mass, respectively.
From \cref{tab:sr}, all the parameters used in the simulations can be retrieved:
\begin{align*}
	\eta_0       &= \dfrac{\rho_p a_0^2 \overline{\dot{\gamma}}}{St};\\
	k_n          &= \hat{k} \eta_0 a_0 \overline{\dot{\gamma}};\\
	k_t          &= \dfrac{2}{7}k_n;\\
	\gamma_n     &= 10^{-7} \dfrac{k_n}{\overline{\dot{\gamma}}};\\
	F^E          &= (6\pi) \eta_0 a_0^2 \dot{\gamma}_0;\\
	F^A          &= \dfrac{F^E}{10}\quad \text{therefore, Hamaker constant};\\
	F^R          &= 9 F^A\quad \text{therefore, $F_0$}.\\
\end{align*}
Note that, for $F^E$ we multiply the non-dimensional group for the value $6\pi$, that
is the value that premultiplies the viscous force in the Stokes drag of a sphere 
\citep[see][where the Stokes drag of a sphere has been chosen as reference value]{MARI2014}.
\begin{table*}
  \begin{center}
     \scalebox{0.8}{
     \setlength{\tabcolsep}{0.55em}
     \begin{tabular}{c|c|c|c|c|c|c|c|c|c|c|c|c}
         \hline
         \rule{0pt}{3ex}
         $\phi$ & $N$ & $a_2/a_1$ & $V_2/V_1$ 
	     & $St$ & $\hat{k}$ & $k_t/k_n$ & $\gamma_n \overline{\dot{\gamma}}/k_n$ 
	     & $\mu_C$ & $\overline{\dot{\gamma}}/\dot{\gamma}_0$ & $\delta_{lub}/a_0$
         & $L_S/a_0$ & $F^R/F^A$ \\[2mm]
	     $0.50$ & $2^{16}$ & $3$ & $0.25$ & $10^{-3}$ & $4\times 10^5$ & $2/7$ & $10^{-7}$
         & $0.5$ & $\{ 10^{-5},\dots,\infty\}$ & $0.25$ & $0.02$ & $9$ \\ \hline
     \end{tabular}}
      \caption{Set of parameters used for simulating the behaviour of
               dense suspensions varying the shear rate. 
               Here, $a_0 = a_1$ is the reference length scale.}
     \label{tab:sr}
   \end{center}
\end{table*}

\section*{Supplementary material to: Results}
\subsection*{Symmetry of the local volume fraction profiles}

\begin{figure*}[t]
\begin{tikzpicture}[thick,scale=0.95, every node/.style={scale=0.95}]
    \begin{axis}[
        hide axis,
        scale only axis,
        height=0pt,
        width=0pt,
        colormap={reddish}{rgb255=(0,0,0) rgb255=(255,215,0) rgb255=(255,128,0) rgb255=(255,  0,  0) rgb255=(153,  0,  0)},
        colorbar sampled,
        colormap access=piecewise const, 
        colorbar,
        point meta min=0,
        point meta max=1,
        colorbar style={
            samples=6,
            height=4cm,
            width=0.3cm,
            ytick style={
                draw=none},
            yticklabels=\empty
            },
    ]
        \addplot [draw=none] coordinates {(0,0)};
    \end{axis}
    \node[text width=1.3cm] at (0.7, 0.3) {$\quad A$};
    \node[text width=1.3cm] at (0.0,-0.4) {$\quad40$};
    \node[text width=1.3cm] at (0.0,-1.2) {$\quad20$};
    \node[text width=1.3cm] at (0.0,-2.0) {$\quad10$};
    \node[text width=1.3cm] at (0.0,-2.8) {$\quad5$};
    \node[text width=1.3cm] at (0.0,-3.6) {$\quad0$};
    \node at (3.7,-2)  {\subfloat[]{\includegraphics[width=0.33\textwidth]{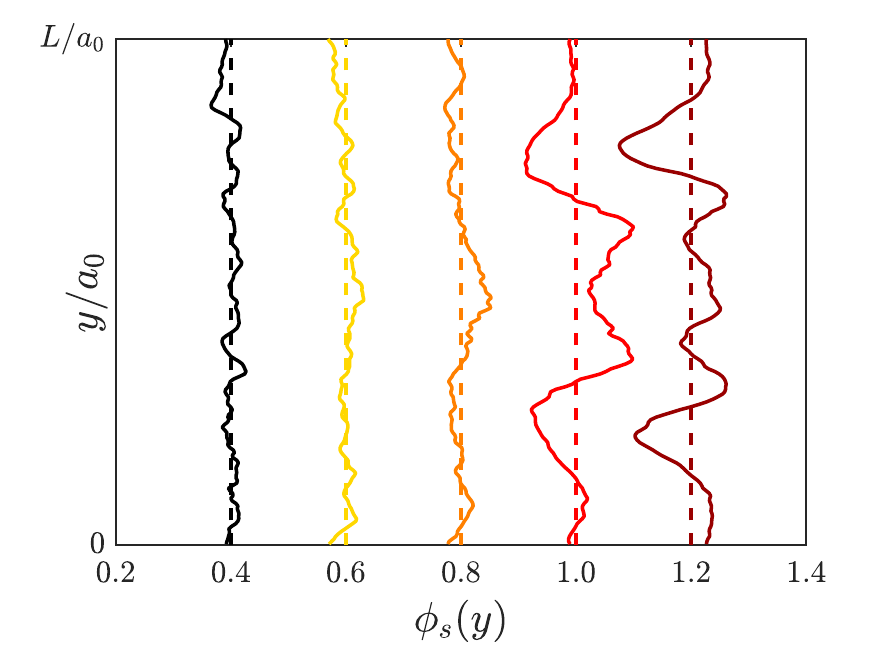}}};%
    \node at (9.2,-2)  {\subfloat[]{\includegraphics[width=0.33\textwidth]{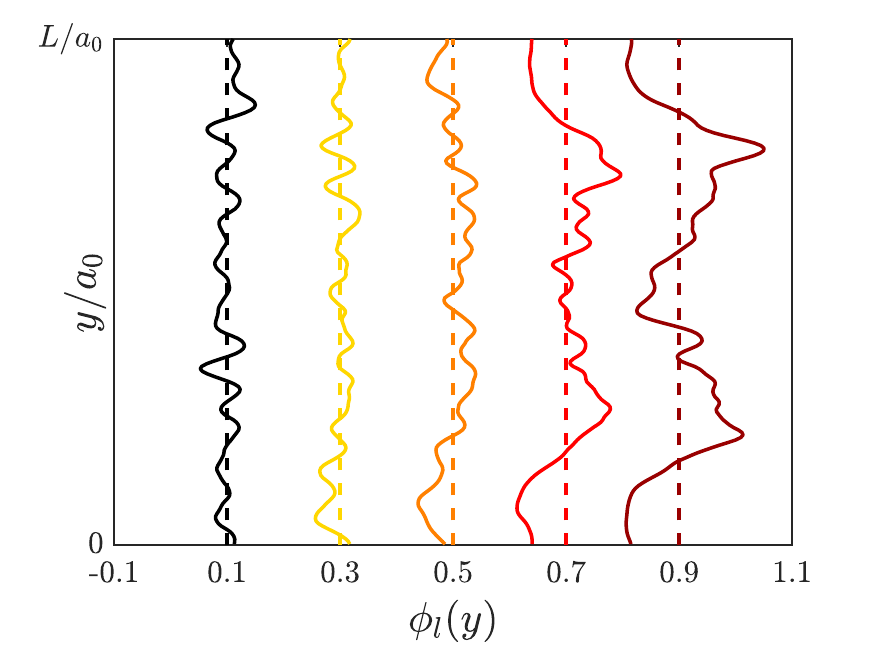}}};%
    \node at (14.7,-2) {\subfloat[]{\includegraphics[width=0.33\textwidth]{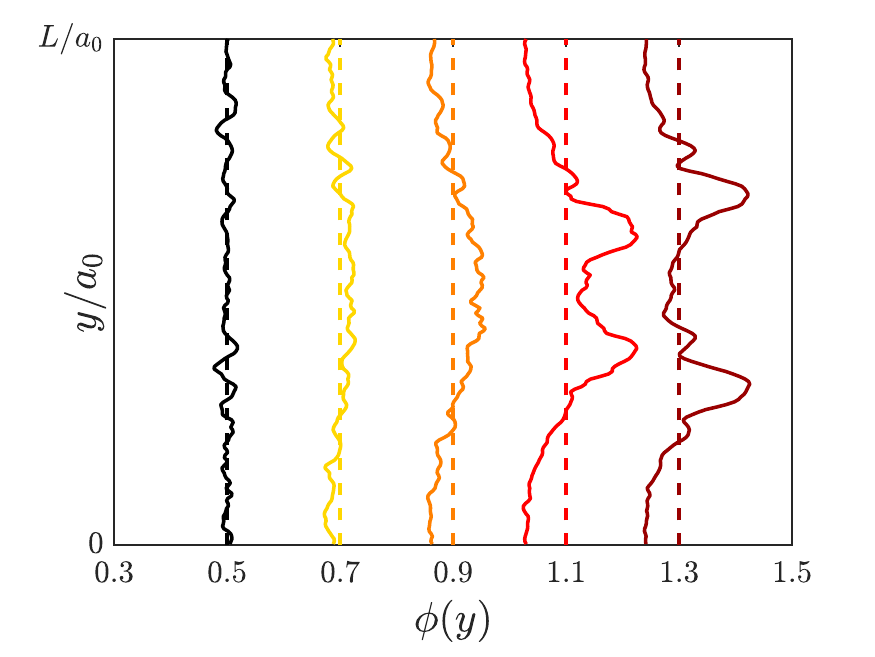}}};%
\end{tikzpicture}\\
\begin{tikzpicture}[thick,scale=0.95, every node/.style={scale=0.95}]
    \begin{axis}[
        hide axis,
        scale only axis,
        height=0pt,
        width=0pt,
        colormap={pinkish}{rgb255=(0,0,0) rgb255=(255,0,0) rgb255=(255,178,204) rgb255=(255,102,153)  rgb255=(178,102,255) rgb255=(127,0,255)},
        colorbar sampled,
        colormap access=piecewise const, 
        colorbar,
        point meta min=0,
        point meta max=1,
        colorbar style={
            samples=7,
            height=4cm,
            width=0.3cm,
            ytick style={
                draw=none},
            yticklabels=\empty
            },
    ]
        \addplot [draw=none] coordinates {(0,0)};
    \end{axis}
    \node[text width=1.3cm] at (0.7, 0.3) {$\quad n$};
    \node[text width=1.3cm] at (0,-0.350) {$\quad16$};
    \node[text width=1.3cm] at (0,-1.025) {$\quad8$};
    \node[text width=1.3cm] at (0,-1.700) {$\quad4$};
    \node[text width=1.3cm] at (0,-2.375) {$\quad2$};
    \node[text width=1.3cm] at (0,-3.050) {$\quad1$};
    \node[text width=1.3cm] at (0,-3.700) {$\quad0$};
    \node at (3.7,-2)  {\subfloat[]{\includegraphics[width=0.33\textwidth]{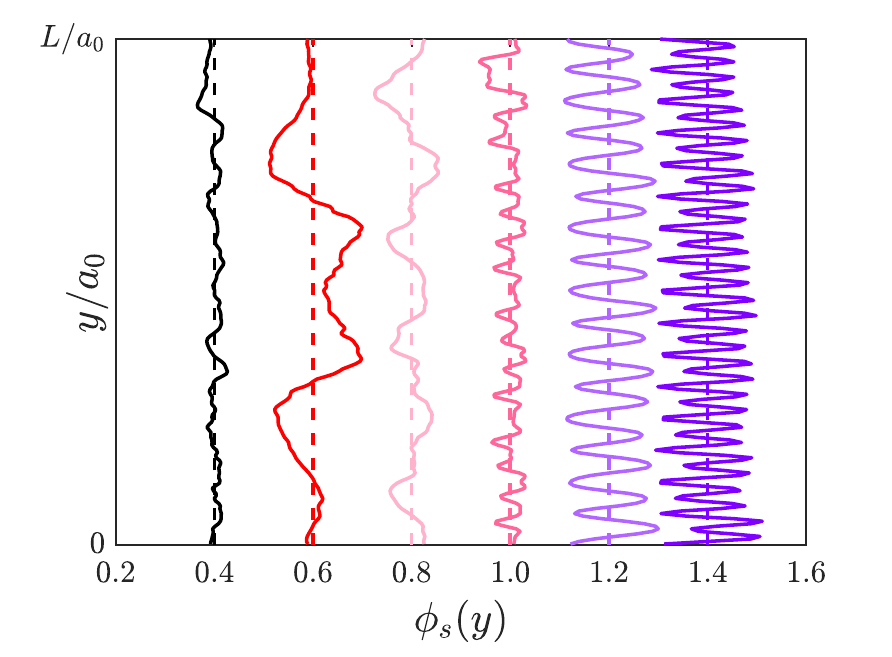}}};%
    \node at (9.2,-2)  {\subfloat[]{\includegraphics[width=0.33\textwidth]{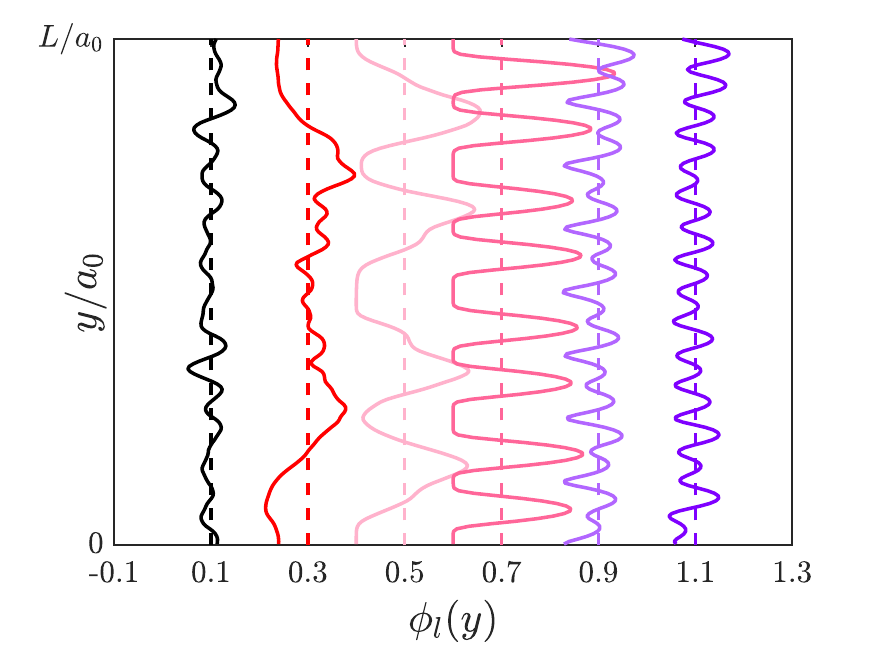}}};%
    \node at (14.7,-2) {\subfloat[]{\includegraphics[width=0.33\textwidth]{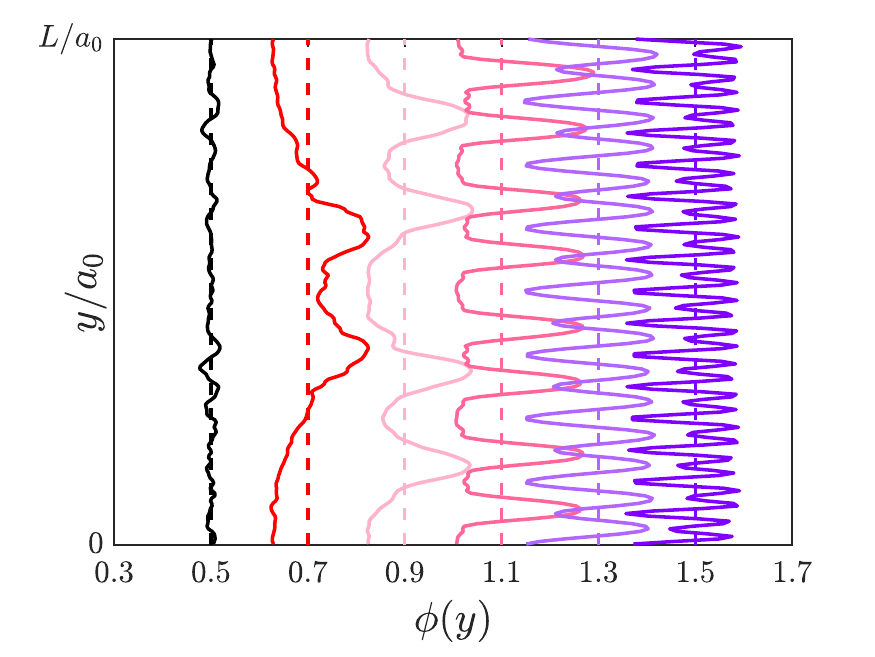}}};%
\end{tikzpicture}\\
\caption{\label{fig:localPhi_old} Profiles of the local volume fractions along the
shearwise direction $y$. Panel \textbf{a} and \textbf{d} show the local volume of 
the small particles $\phi_s(y)$; panel \textbf{b} and \textbf{e} show the local 
volume of the large particles $\phi_l(y)$; panel \textbf{c} and \textbf{f} show 
the total local volume fraction $\phi(y)$. The panels in the top row, i.e.\ 
panel \textbf{a}--\textbf{c}, refer to the sinusoidal
disturbance with $n=1$ and increasing amplitude $A$, as shown in the respective
legend on the left of the figure; the panels in the bottom row, i.e.\ panel 
\textbf{d}--\textbf{f}, refer to the cases with $A=20$ and increasing $n$, as shown in the 
respective legend on the left of the figure. The black profiles in each panel
refer to the reference case with no sinusoidal disturbance. Finally, the dashed
lines refer to the values $\overline{\phi}_s=0.4$, $\overline{\phi}_l=0.1$ and 
$\overline{\phi}=0.5$. Note that, the profiles for different $A$ and $n$ have been 
shifted horizontally by $0.2$ for visual purpose. Version with no ensemble average.}

\end{figure*}

\begin{figure*}[t]
\begin{tikzpicture}[thick,scale=0.95, every node/.style={scale=0.95}]
    \begin{axis}[
        hide axis,
        scale only axis,
        height=0pt,
        width=0pt,
        colormap={reddish}{rgb255=(0,0,0) rgb255=(255,215,0) rgb255=(255,128,0) rgb255=(255,  0,  0) rgb255=(153,  0,  0)},
        colorbar sampled,
        colormap access=piecewise const, 
        colorbar,
        point meta min=0,
        point meta max=1,
        colorbar style={
            samples=6,
            height=4cm,
            width=0.3cm,
            ytick style={
                draw=none},
            yticklabels=\empty
            },
    ]
        \addplot [draw=none] coordinates {(0,0)};
    \end{axis}
    \node[text width=1.3cm] at (0.7, 0.3) {$\quad A$};
    \node[text width=1.3cm] at (0.0,-0.4) {$\quad40$};
    \node[text width=1.3cm] at (0.0,-1.2) {$\quad20$};
    \node[text width=1.3cm] at (0.0,-2.0) {$\quad10$};
    \node[text width=1.3cm] at (0.0,-2.8) {$\quad5$};
    \node[text width=1.3cm] at (0.0,-3.6) {$\quad0$};
    \node at (3.7,-2)  {\subfloat[]{\includegraphics[width=0.33\textwidth]{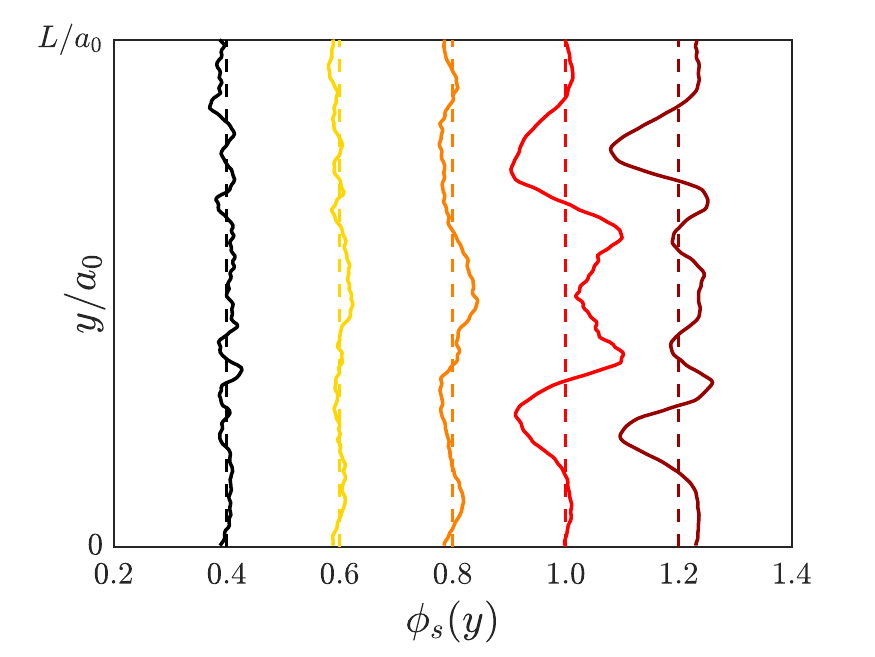}}};%
    \node at (9.2,-2)  {\subfloat[]{\includegraphics[width=0.33\textwidth]{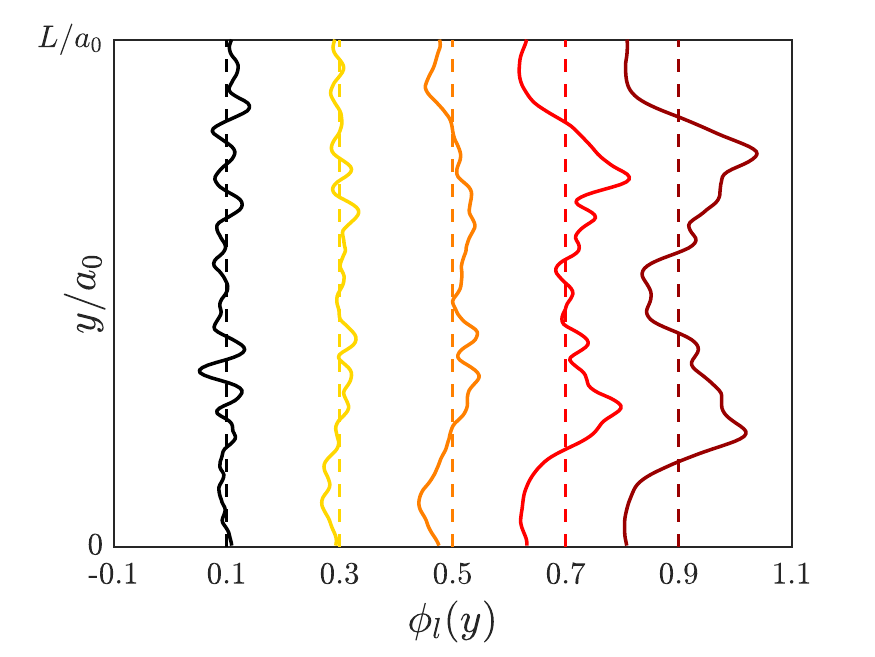}}};%
    \node at (14.7,-2) {\subfloat[]{\includegraphics[width=0.33\textwidth]{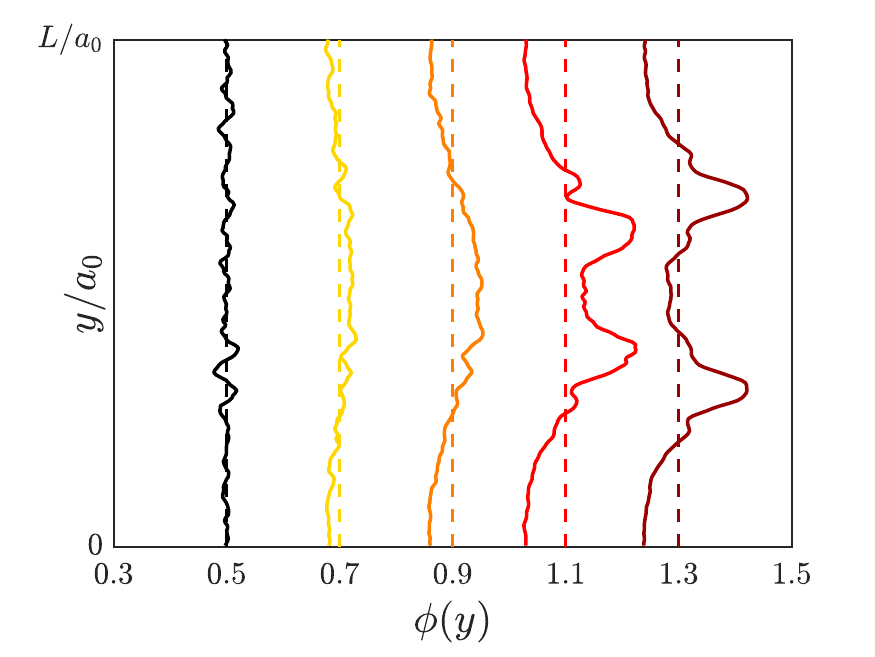}}};%
\end{tikzpicture}\\
\begin{tikzpicture}[thick,scale=0.95, every node/.style={scale=0.95}]
    \begin{axis}[
        hide axis,
        scale only axis,
        height=0pt,
        width=0pt,
        colormap={pinkish}{rgb255=(0,0,0) rgb255=(255,0,0) rgb255=(255,178,204) rgb255=(255,102,153)  rgb255=(178,102,255) rgb255=(127,0,255)},
        colorbar sampled,
        colormap access=piecewise const, 
        colorbar,
        point meta min=0,
        point meta max=1,
        colorbar style={
            samples=7,
            height=4cm,
            width=0.3cm,
            ytick style={
                draw=none},
            yticklabels=\empty
            },
    ]
        \addplot [draw=none] coordinates {(0,0)};
    \end{axis}
    \node[text width=1.3cm] at (0.7, 0.3) {$\quad n$};
    \node[text width=1.3cm] at (0,-0.350) {$\quad16$};
    \node[text width=1.3cm] at (0,-1.025) {$\quad8$};
    \node[text width=1.3cm] at (0,-1.700) {$\quad4$};
    \node[text width=1.3cm] at (0,-2.375) {$\quad2$};
    \node[text width=1.3cm] at (0,-3.050) {$\quad1$};
    \node[text width=1.3cm] at (0,-3.700) {$\quad0$};
    \node at (3.7,-2)  {\subfloat[]{\includegraphics[width=0.33\textwidth]{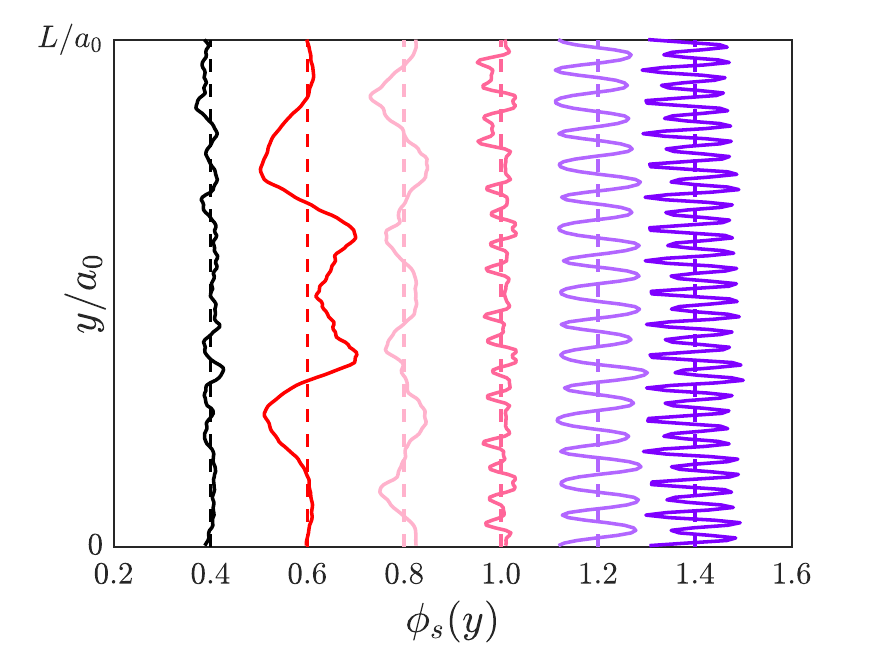}}};%
    \node at (9.2,-2)  {\subfloat[]{\includegraphics[width=0.33\textwidth]{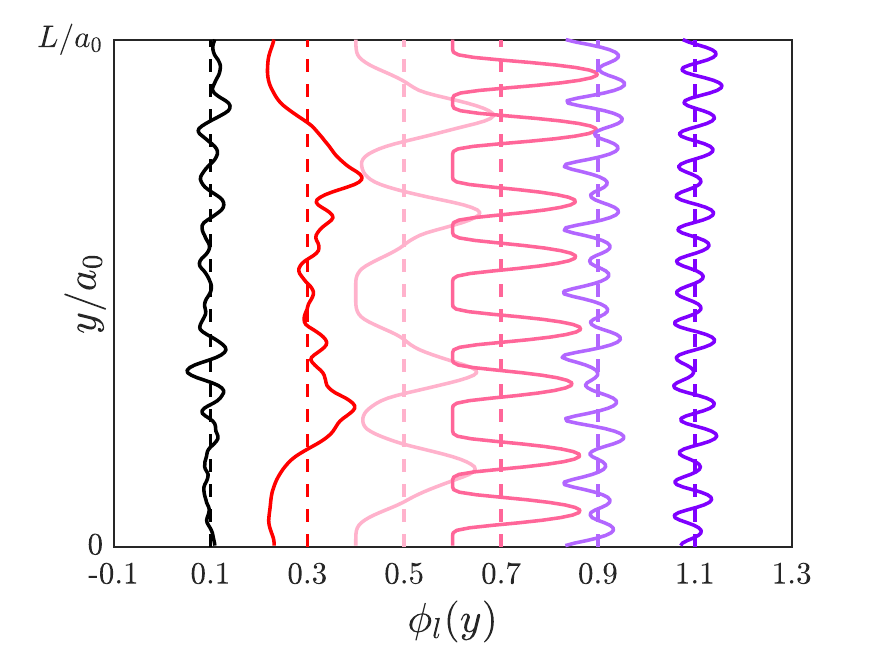}}};%
    \node at (14.7,-2) {\subfloat[]{\includegraphics[width=0.33\textwidth]{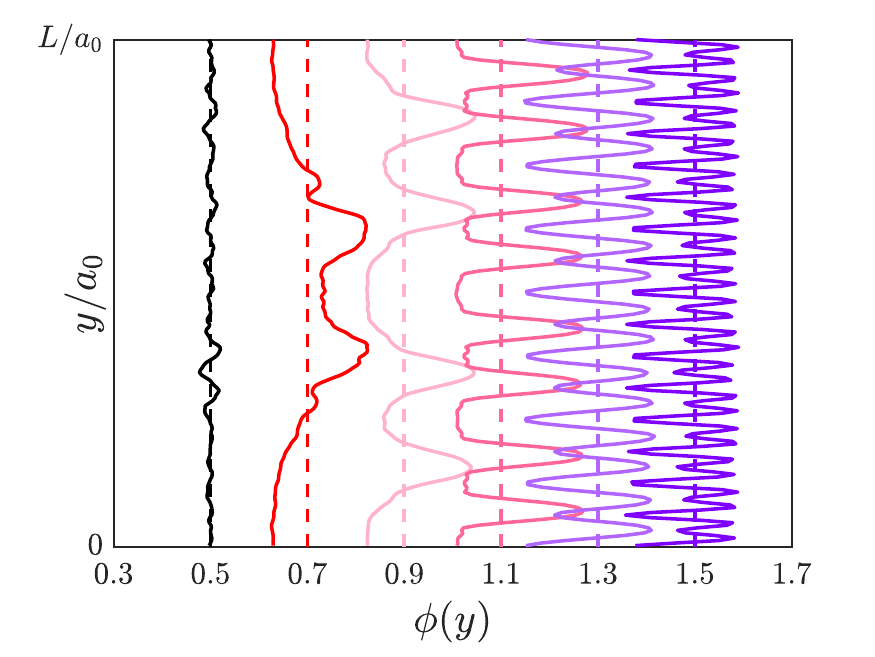}}};%
\end{tikzpicture}\\
\caption{\label{fig:localPhi_new} Profiles of the local volume fractions along the
shearwise direction $y$. Panel \textbf{a} and \textbf{d} show the local volume of 
the small particles $\phi_s(y)$; panel \textbf{b} and \textbf{e} show the local 
volume of the large particles $\phi_l(y)$; panel \textbf{c} and \textbf{f} show 
the total local volume fraction $\phi(y)$. The panels in the top row, i.e.\ 
panel \textbf{a}--\textbf{c}, refer to the sinusoidal
disturbance with $n=1$ and increasing amplitude $A$, as shown in the respective
legend on the left of the figure; the panels in the bottom row, i.e.\ panel 
\textbf{d}--\textbf{f}, refer to the cases with $A=20$ and increasing $n$, as shown in the 
respective legend on the left of the figure. The black profiles in each panel
refer to the reference case with no sinusoidal disturbance. Finally, the dashed
lines refer to the values $\overline{\phi}_s=0.4$, $\overline{\phi}_l=0.1$ and 
$\overline{\phi}=0.5$. Note that, the profiles for different $A$ and $n$ have been 
shifted horizontally by $0.2$ for visual purpose. Version with ensemble average.}
\end{figure*}

Concerning the possible lack of symmetry of the profiles in Figure 4 of the main manuscript, we prove here that the lack of averaging and the 
finite number of particles (only 600 large particles) are to blame.
In particular, 
we considered three initial conditions to the 8 cases with the sinusoidal perturbation, obtained by sampling the simulation with no perturbation 
(i.e. $A=0$ and $n=0$) at different time instants.
The cases with identical amplitude and wavenumber are then ensemble averaged, together 
with 900 instantaneous fields per run sampled after convergence. 
\Cref{fig:localPhi_old,fig:localPhi_new} compare the profiles with and without ensemble average
respectively. The improvement of the symmetry is clearly visible.
Therefore, the lack of perfect symmetry in the volume faction profiles is related 
to the lack of averaging only, together with the finite number of particles.

Finally, we also prove the correctness of our implementation of the Lees-Edwards boundary conditions.
We selected the run with $A=20$ and $n=1$ (red curve in \cref{fig:localPhi_old,fig:localPhi_new})
and we averaged its volume fraction distributions with an identical set-up whose
streamwise velocity profile is shifted of a value $u = 30 a_0 \overline{\dot{\gamma}}$.
In case of a wrong implementation of the boundary conditions, the resulting local
volume fraction should differ, while the red profiles in \cref{fig:localPhi_new} demonstrate
the correctness of the implementation.

\bibliography{biblio}